\documentclass{aa}  
\newcommand{\PSF}{\textrm{psf}}
\usepackage{graphicx}

\makeatletter
\renewcommand*\aa@pageof{, page \thepage{} of \pageref*{LastPage}}
\makeatother

\usepackage{txfonts}

\usepackage[colorlinks,linkcolor={blue},citecolor={blue},urlcolor={blue}]{hyperref}

\usepackage{mathrsfs}
\usepackage[normalem]{ulem}

\defcitealias{samuroff_dark_2023}{Sa23}
\newcommand{\Sam}{\citetalias{samuroff_dark_2023}}

\defcitealias{zhang_point-spread_2024}{Z24}
\newcommand{\ziwen}{\citetalias{zhang_point-spread_2024}}

\defcitealias{singh_intrinsic_2015}{Si15}
\newcommand{\Singh}{\citetalias{singh_intrinsic_2015}}

\begin{document}

   \title{UNIONS: a direct measurement of intrinsic alignment with BOSS/eBOSS spectroscopy}

   \author{%
          Fabian Hervas Peters\inst{1}
          \fnmsep\thanks{fabian.hervaspeters@cea.fr}
          \and
          Martin Kilbinger
          \inst{1}          
          \and
          Romain Paviot \inst{1}
          \and
          Lucie Baumont \inst{1,2,3}
          \and
          Elisa Russier \inst{1,4}
          \and
          Ziwen Zhang \inst{1,5,6} 
          \and
          Calum Murray \inst{1}
          \and
          Valeria Pettorino \inst{1,7}
          \and
          Thomas de Boer \inst{8}
          \and
          S\'ebastien Fabbro \inst{9}
          \and
          Sacha Guerrini \inst{1}
          \and
          Hendrik Hildebrandt \inst{10}
          \and
          Mike Hudson \inst{11,12,13}
          \and
          Ludovic Van Waerbeke \inst{14}
          \and
          Anna Wittje \inst{10}
    }
   \institute{
        Université Paris-Saclay, Université Paris Cité, CEA, CNRS, AIM, 91191, Gif-sur-Yvette, France
        \and
        Dipartimento di Fisica - Sezione di Astronomia, Università di Trieste, Via Tiepolo 11, 34131 Trieste, Italy
        \and
        INAF-Osservatorio Astronomico di Trieste, Via G. B. Tiepolo 11, 34143 Trieste, Italy
        \and
        Lawrence Berkeley National Laboratory, 1 Cyclotron Road, Berkeley, CA 94720, USA
        \and
        CAS Key Laboratory for Research in Galaxies and Cosmology, Department of Astronomy, University of Science and Technology of China, Hefei, Anhui 230026, China
       \and
       School of Astronomy and Space Science, University of Science and Technology of China, Hefei 230026, China
        \and
        European Space Agency/ESTEC, Keplerlaan 1, 2201 AZ Noordwijk, The Netherlands
        \and
        Institute for Astronomy, University of Hawaii, 2680 Woodlawn Drive, Honolulu HI 96822
        \and
        NRC Herzberg Astronomy \& Astrophysics, 5071 West Saanich Road, British Columbia, Canada V9E2E7
        \and
        Ruhr University Bochum, Faculty of Physics and Astronomy, Astronomical Institute (AIRUB), German Centre for Cosmological Lensing, 44780 Bochum, Germany
        \and
        Waterloo Centre for Astrophysics, University of Waterloo, Waterloo, ON N2L 3G1, Canada
        \and 
        Department of Physics and Astronomy, University of Waterloo, Waterloo, ON N2L 3G1, Canada
        \and
        Perimeter Institute for Theoretical Physics, Waterloo, ON N2L 2Y5, Canada
        \and
         Department of Physics and Astronomy, University of British Columbia, Vancouver, V6T1Z1, BC, Canada
    }

   \date{Received XXX, ; accepted YYY}

\abstract
{
 During their formation, galaxies are subject to tidal forces, which create correlations between their shapes and the large-scale structure of the Universe, known as intrinsic alignment. This alignment is a contamination for cosmic-shear measurements as one needs to disentangle correlations induced by external lensing effects from those intrinsically present in galaxies. 
 }
{
We constrain the amplitude of intrinsic alignment and test models by making use of the overlap between the Ultraviolet Near-Infrared Optical Northern Survey (UNIONS) covering $3,500 \deg^2$, and spectroscopic data from the Baryon Oscillation Spectroscopic Survey (BOSS/eBOSS). By comparing our results to measurements from other lensing surveys on the same spectroscopic tracers, we can test the reliability of these estimates and verify they are not survey dependent. 
}
{ We measure projected correlation functions between positions and ellipticities, which we model with perturbation theory to constrain the commonly used non-linear alignment model and its higher-order expansion. We compute an analytical covariance matrix which we validate with jackknife estimates.
}
{ 
Using the non-linear alignment model, we obtain a $13\sigma$ detection with CMASS galaxies, a $3\sigma$ detection with LRGs, and a detection compatible with the null hypothesis for ELGs. We test the tidal alignment and tidal torque model, a higher-order alignment model, which we find to be in good agreement with the non-linear alignment prediction and for which we can constrain the second-order parameters. We show a strong scaling of our intrinsic alignment amplitude with luminosity. We demonstrate that the UNIONS sample is robust against systematic contributions, particularly concerning PSF biases. We reached a reasonable agreement when comparing our measurements to other lensing samples for the same spectroscopic samples. We take this agreement as an indication that direct measurements of intrinsic alignment are mature for stage IV priors. 
}
{}


   \maketitle
%

\section{Introduction}

Cosmic shear measurements have demonstrated in recent years their capacity to constrain the dark-matter distribution on cosmological scales \citep{dark_energy_survey_and_kilo-degree_survey_collaboration_y3_2023, more_hyper_2023}. They represent a key probe in the coming decade to determine our Universe's cosmological scenario in later times and to gain insights into the nature of dark energy and dark matter at lower and intermediate redshift. Weak lensing is one of the primary cosmological probes for a number of upcoming surveys, including \textit{Euclid} \citep{euclid_collaboration_euclid_2024}, the Vera Rubin Observatory Large Survey of Space and Time \citep[LSST;][]{ivezic_lsst_2018} and the Nancy Grace Roman Space Telescope \citep{eifler_cosmology_2021}. The cosmic shear signal is estimated by averaging over the shape of background galaxies, which are coherently deformed as their light travels through the large scale structures. By studying the correlations between the shapes of neighboring galaxies, one can infer the foreground distribution of matter, particularly dark matter. In the coming decade, unprecedented statistical power will be reached by modern weak lensing experiments, which will observe billions of galaxies. 
This increase in statistical power means that systematic contributions, which were previously sub-dominant, can lead to important biases if not accounted for. These systematic contributions can be broadly split into two categories: biases at the measurement level, and biases at the modelling level. In the latter, two sources of biases are particularly critical: baryonic feedback, a non-linear effect that suppresses large $k$ mode fluctuations, and galaxy intrinsic alignment, as galaxy orientations are not uniformly distributed on the sky, but are correlated with their environments. For a review, see \citet{mandelbaum_weak_2018}.

It was shown in \citet{catelan_intrinsic_2001} and \citet{crittenden_spin-induced_2001} that tidal forces arising from the density field have the potential to produce coherent orientations in neighboring galaxies, which can be a challenge for cosmic shear: The difficulty is to disentangle the correlations
induced directly on the galaxy shapes by the tidal forces from those caused by the shearing of the light profile due to gravitational lensing. The effects of tidal forces on the orientation of galaxies inside haloes have been described in \citet{catelan_intrinsic_2001} and in \citet{mackey_theoretical_2002}. For a review and visual explanations, see \citet{joachimi_galaxy_2015,lamman_ia_2024}.

To mitigate intrinsic alignment, various solutions have been proposed. To remove intrinsic alignment from the cosmic-shear signal, two main methods exist. These are ``down-weighting'' \citep{king_separating_2003} and ``nulling'' \citep{heavens_cosmic_2011}. Down-weighting consists of suppressing the correlation between pairs of galaxies close in redshift, which are subject to intrinsic alignment. This is achieved by applying a Gaussian kernel in the correlation function estimator. This kernel has a variance reflecting the photometric redshift uncertainty, which implies that imprecise photometric redshifts will lead to a large loss of statistical power. The concept of ``nulling'' uses the specific dependency of the lensing kernel on the comoving distance to suppress any combination of bins too sensitive to the interplay between shear and intrinsic alignment. For a review, see \citet{kirk_galaxy_2015}.

These methods, however, have been shown to strongly reduce the statistical power of a weak lensing survey, and to be very dependent on precise photometric redshifts. 
The approach taken in recent stage-III cosmic shear surveys has therefore been the joint modeling of the weak lensing and the intrinsic alignment signal. Two main models have been applied, the non-linear alignment model (NLA), and its higher-order expansion, the tidal alignment and tidal torque model (TATT). These models are described in Sects.~\ref{sec:NLA} and \ref{sec:TATT}. They are theoretically motivated, but the physical range of their validity remains an open question. In particular, \citet{bakx_effective_2023} have shown recently that the NLA (TATT) prediction of the alignment of subhaloes in the DarkQuest simulation can only describe correlations on modes smaller than $k \leq 0.08 \, h/$Mpc ($k\leq0.15 \, h$/Mpc). This shortcoming motivated the inclusion of even higher order terms in the effective field theory approach. With 8 free parameters \citep{vlah_eft_2020}, this model is harder to constrain than the NLA and TATT models, which have 1 and 3 parameters, respectively. Note that while these theoretically motivated models are essential in the prediction of the weak lensing correlation functions, they are also used to add intrinsic alignment to convergence maps used in simulation-based approaches, which require the full-field information \citep{harnois-deraps_cosmic_2021,kacprzak_cosmogridv1_2022,jeffrey_dark_2024}.
Other models based on the halo model \citep{schneider_halo_2010,fortuna_halo_2021} are accurate down to smaller scales by modeling a strong intrinsic ellipticity field inside each halo.

To properly characterize intrinsic alignment, various hydrodynamical simulations have been used \citep{chisari_intrinsic_2015,samuroff_advances_2021}. While they show a picture broadly consistent with observations, no precise calibration was reached from these various simulations as the selection of galaxies and their mapping to phenomenological properties remains challenging. In \cite{tenneti_intrinsic_2016}, it was shown that the choice of elliptical galaxies can produce drastic changes as the ellipticity-density correlation can go from positively to negatively correlated depending on the simulation. 
A newer route to calibrate intrinsic alignment, which makes use of the power of $N$-body simulations to describe the large-scale structure of the Universe, is to add galaxies empirically. This is performed by drawing misalignment angles from probability distribution functions, which are calibrated on observations or hydrodynamical simulations \citep{hoffmann_modeling_2022,van_alfen_empirical_2024}.

In parallel, and complementary to the simulation efforts, many studies have tried to constrain intrinsic alignment models by direct measurements, which require precise redshift information to disentangle lensing contributions from intrinsic alignment. They have shown a strong intrinsic alignment between red galaxies at scales relevant for the 2-halo term (i.e., $\sim 6$--$150$ Mpc) outside the 1-halo regime  (\citealt[][{[}hereafter \Singh{]}]{joachimi_intrinsic_2013,singh_intrinsic_2015}; \citealt{johnston_kidsgama_2019}; \citealt{fortuna_kids-1000_2021}; \citealt{samuroff_dark_2018}).  For red galaxies, intrinsic alignment was shown to depend on luminosity following a broken power law, (\citealt[][{[}hereafter \Sam{]}]{fortuna_halo_2021,samuroff_dark_2023}). For blue spiral galaxies, direct measurements tend to prefer an absence of intrinsic alignment 
(\citealt{mandelbaum_wigglez_2011}; \citealt{johnston_kidsgama_2019}; \citealt{tonegawa_intrinsic_2024}; \Sam) . The alignment between blue and red samples was well measured in \citet{johnston_kidsgama_2019} which was made possible by the very densely sampled and complete spectroscopic galaxy and mass assembly survey \citep[GAMA,][]{hopkins_galaxy_2013}, overlapping with the KiDS dataset \citep{giblin_kids-1000_2021}. Very recently, \citet{mccullough_dark_2024} showed a strong difference in cosmological parameter estimation from cosmic shear when splitting the DES Y3 in red and blue galaxy samples. This difference was attributed to the varying sensitivities of these samples to the intrinsic alignment models.

This work contributes to the challenge of propagating intrinsic alignment measurements into a sensible prior for stage-IV surveys. This task is complicated by a few subtleties affecting observations and the resulting amplitude of intrinsic alignment. One such effect is the dependence of the intrinsic alignment amplitude on the shape measurement method: As tidal forces act stronger on the outskirts of galaxies, these extended regions are more affected by intrinsic alignment than the central part. This was first observed in \citet{singh_intrinsic_2016}, who found a 40 percent amplitude difference between the re-Gaussianization, isophotal, and de-Vaucouleurs shape measurement methods. While these three shape measurement methods are less used in stage-III surveys, recent work \citep{macmahon_intrinsic_2023} comparing \texttt{Metacalibration} and \texttt{im3shape} also predict a potential 40 percent difference for LSST Y1 data. In their work, they apply the mean estimator method (MEM) described in \cite{leonard_measuring_2018}, which uses the different deformations in isophotes between intrinsic alignment and cosmic shear to disentangle the two effects.  In \cite{georgiou_dependence_2019,georgiou_gamakids_2019}, a strong change in IA amplitude was shown by varying the weight function of the shape measurement. In addition \cite{georgiou_dependence_2019} noted a chromatic effect in intrinsic alignment by changing the photometric band in which the galaxy shape is measured. 

In this work, we apply the projected two-point correlation function, but other estimators have recently been developed. In \cite{linke_third-order_2024}, a three-point correlation function with LOWZ data was found to be in good agreement with the two-point correlation functions, promising better constraining power for future surveys. Another important avenue is the 3-dimensional measurement in Fourier space, known as the Yamamoto estimator \citep{yamamoto_measurement_2006}, which was initially used by the galaxy clustering community but has been applied successfully to constrain intrinsic alignment \citep{kurita_power_2021,kurita_constraints_2023}. Similarly, \citet{singh_increasing_2024} showed that including real-space multipoles can also help in reaching tighter constraints.

Whereas intrinsic alignment is a contamination to weak lensing measurements, it is also a cosmological probe in its own right. Examples include the detection of baryon acoustic oscillations in the intrinsic-alignment correlation functions \citep{van_dompseler_alignment_2023,xu_evidence_2023}, and constraints on primordial non-Gaussianities \citep{schmidt_imprint_2015,kurita_constraints_2023}.

In this paper, we use the UNIONS imaging survey together with BOSS/eBOSS spectroscopic galaxies to measure intrinsic alignment. We compare our measurements to other samples by investigating the dependence on luminosity before taking a deeper look at intrinsic alignment measurements from SDSS lensing and DES Y3 lensing on the same set of galaxies. We verify that our measurements are robust against systematic diagnostics, in particular with respect to PSF multiplicative and additive biases. 

This paper is organized as follows:
We start in Sect.~\ref{sec:cat} by presenting the datasets used to constrain intrinsic alignment models. In Sect.~\ref{sec:corr}, we 
we present the various estimators correlating shapes and positions.
Sect.~\ref{sec:model} introduces the NLA and TATT models and the calculations of the predictions for the estimators. We also present the calculation of the analytical covariance matrices, which we validate with jackknife estimates.
We show our results in Sect.~\ref{sec:results},
and conclude in Sect.~\ref{sec:conclusion}.

\section{Data} \label{sec:cat}

\subsection{UNIONS catalogues}\label{UNIONS}
We use  $3,500$ deg$^2$ of imaging data from  UNIONS, the Ultraviolet Near-Infrared Optical Northern Survey. UNIONS groups deep-field observations obtained by the CFHT (Canada-France-Hawai'i Telescope on Mauna Kea), Pan-STARRS (Panoramic Survey Telescope and Rapid Response System on Maui), and Subaru (on Mauna Kea) telescopes in the Northern hemisphere. The final observations will contain multi-band ($ugriz$) information to provide precise photometric redshifts for \textit{Euclid}. In this work we use shape measurements realized in the $r$ band observed by CFHT. The astrometry and image reduction were obtained with the Megapipe pipeline \citep{gwyn_megapipe_2008}. Through this work, we use a catalogue created with the {\fontfamily{lmtt}\selectfont ShapePipe}\footnote{\color{blue}{\url{https://github.com/CosmoStat/shapepipe}}} pipeline \citep{farrens_shapepipe_2022}. A first realisation of the catalogue covering $1,500$ deg$^2$ and systematics diagnostics are described in \citet{guinot_shapepipe_2022}. The {\fontfamily{lmtt}\selectfont Shapepipe} pipeline uses the \texttt{ngmix} \citep{2015ascl.soft08008S} software for shape measurement, with a 2D Gaussian profile to fit the galaxy light distribution, and \texttt{Metacalibration} \citep{sheldon_practical_2017} for shear calibration. \texttt{Metacalibration} is a framework in which galaxies are artificially sheared during the shape measurement process to evaluate a finite  $\partial \varepsilon\ / \partial \gamma$ correction. This gives a shear response $R$, evaluated between positively and negatively sheared postage stamps via finite differences. This artificial shearing makes it possible to quantify model, noise and selection biases by evaluating how the estimated ellipticity varies when it is changed by a known shear. Objects are detected and deblended with the SExtractor software \citep{1996A&AS..117..393B}. In this analysis, we included objects with \texttt{FLAGS}$\leq 2$ from SExtractor as we are working with spectroscopically confirmed galaxies only. 
The catalogue comes with inverse variance weights related to errors on the shape measurements given by \texttt{ngmix}, 
\begin{equation}
    w_{\mathrm{SP}}=\frac{1}{2 \sigma^2_{\mathrm{SN}}+\sigma^2_{e_1}+\sigma^2_{e_2}}
\end{equation}
where $\sigma_{\mathrm{SN}}$ is the galaxy shape noise and $\sigma_{e_i}$ represents the variance of the ellipticity component $i=1, 2$ estimated by the model fitting procedure.

The PSF is modeled with {\fontfamily{lmtt}\selectfont MCCD}\footnote{\url{https://github.com/CosmoStat/mccd}}  \citep{liaudat_multi-ccd_2021}, a software based on a minimization scheme analogous to {\fontfamily{lmtt}\selectfont PSFEx} but which includes both a \textit{local} component fitted on a single CCD and a \textit{global} contribution, which captures effects varying across the entire focal plane.

\begin{figure}
    \centering
    \includegraphics[width=9cm]{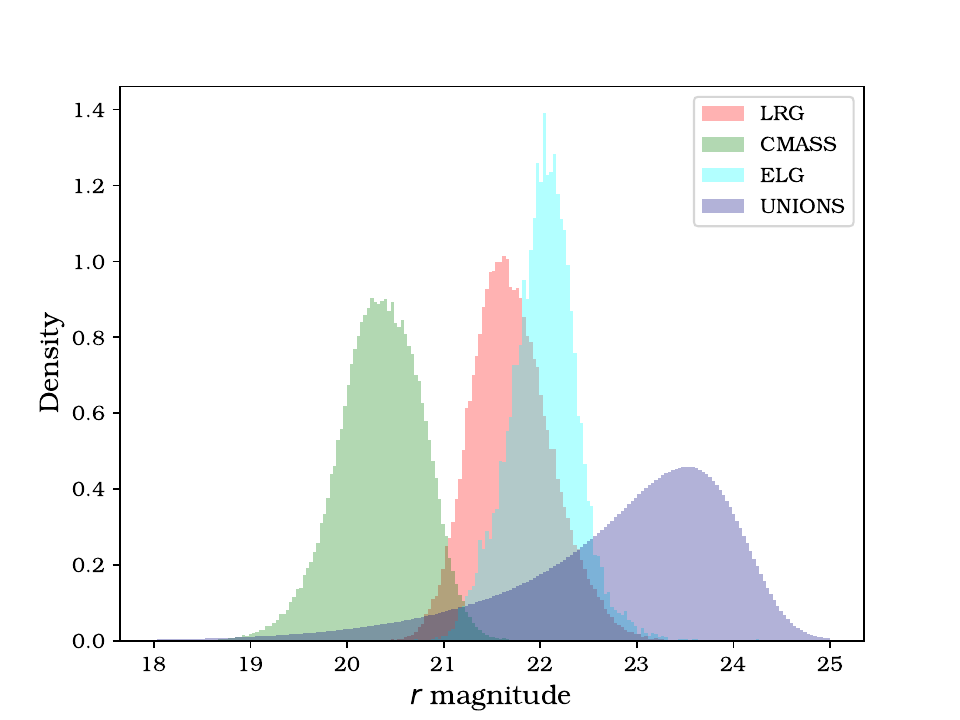}
    \caption{The unweighted normalized $r$-band magnitude distributions of the BOSS/eBOSS spectroscopic samples and UNIONS photometric galaxies. One can see that the spectroscopic galaxies from the BOSS/eBOSS surveys fall into the the brighter tail of the UNIONS distribution. The galaxies which dominate the lensing signal are comparatively fainter. Note that we could not use our large overlap with LOWZ galaxies as these bright galaxies ($r$<20) are mostly absent from our lensing sample due to size cuts.}
    \label{rmag}
\end{figure}

The objects entering the lensing catalogue are selected with a signal-to-noise ratio (S/N) > 10 , which is estimated from the fitted flux divided by the flux error. The UNIONS catalogue reaches an $r$-magnitude depth of $r\approx24.7$ as shown in Fig.~\ref{rmag}.

For validation, we compare a correlation function measurement obtained with {\tt{ShapePipe}} to one produced with \textit{lensfit} in Appendix.\ref{appen:lensfit}.

\subsection{DES Y3 catalogue}\label{DES}
For a comparison between UNIONS and DES intrinsic alignment measurements, we reproduce Dark Energy Survey measurements close to those presented in \Sam. 
For this purpose, we use the publicly available DES Y3 {\tt{Metacalibration}} shape catalogue. The catalogue is described in \citet{gatti_dark_2021}. It contains $100,204,026$ galaxies covering $4,139$ square degrees of sky area. 
Images were taken in the $g$, $r$, $i$, $z$, and $Y$ bands and galaxy shapes were fitted on $r$, $i$ $z$ band images simultaneously.

The core of the galaxy shear estimation methods is common to DES and UNIONS, which both use the \texttt{ngmix} model fitting software, and calibrate the shear response with the \texttt{Metacalibration} scheme. This is critical for our comparison, as it was shown in \citet{georgiou_dependence_2019} and \citet{singh_intrinsic_2016} that different shape measurement algorithms can lead to very different intrinsic alignment amplitudes.

To create the shear and redshift galaxy sample, we first match the BOSS and eBOSS spectroscopic galaxies to the full DES Y3 source galaxies (see Sect.~\ref{sec:matching}). We then remove spurious objects by applying cuts based on the size and signal-to-noise ratio, and the likelihood of being contaminated by
binary stars as described in \citet{gatti_dark_2021}.

Below we highlight differences between the UNIONS and DES Y3 catalogues and methods which are important in order to compare the measurements from each dataset:
\begin{itemize}
\item DES uses the \textsc{Piff} \citep{jarvis_dark_2021} PSF model while the UNIONS catalogue is built with {\fontfamily{lmtt}\selectfont MCCD}. \textsc{Piff} in the DES Y3 setup fits the PSF CCD by CCD, while {\fontfamily{lmtt}\selectfont MCCD} captures the entire focal plane at once.
\item UNIONS measures galaxy shapes by fitting the CFHT $r$ band, while DES provides joint fits using the $r, i$, and $z$ band images.  
\item The seeing between the two data sets is very different; UNIONS carried out observations with a median seeing of $0.69\arcsec$ in $r$, while DES images have a median seeing of $0.95\arcsec$, $0.88\arcsec$, $0.83\arcsec$ in $r, i$, and $z$, respectively \citep{abbott_dark_2021}. 

\end{itemize}

\subsection{BOSS/eBOSS catalogues} \label{BOSS_cat}

The BOSS/eBOSS tracer catalogues were selected for galaxy clustering studies. Using these catalogues to measure the intrinsic alignment signal is therefore advantageous as these galaxies have been selected by phenomenological properties and have a well-characterized galaxy bias. Such homogeneous galaxy samples are not representative of the complete set of a typical weak-lensing galaxy set, and the corresponding intrinsic alignment measurements cannot be adopted directly as prior for weak lensing. To measure intrinsic alignment, we need to model galaxy bias parameters, over which we marginalize. These parameters, therefore, need to be well constrained to avoid degeneracies with intrinsic alignment parameters.

The catalogue contains weights for each galaxy described in \citet{reid_sdss-iii_2015,raichoor_completed_2020}, which are the product of the following components:
\begin{itemize}
\item $w_\textrm{cp}$, to account for fiber collisions in galaxy pairs;
\item $w_\textrm{noz}$, to account for model redshift failures;
\item $w_\textrm{sys, tot}$, \ to remove spurious observationally induced fluctuations of non-cosmological origin  ;
\item $w_\textrm{FKP}$,
to minimize the mean square fluctuations of the density power spectrum $P_\delta(k)$, introduced in \citet{feldman_power_1994}.
They are defined as $w_\textrm{FKP}(z) = (1 + \bar{n}(z) P_0)^{-1}$ with $P_0$ being the value of the power spectrum at $k\approx 0.15 \, h^{-1}\mathrm{Mpc}$ and $\bar{n}(z)$ the mean galaxy density at redshift $z$.
\end{itemize}

\begin{figure}
    \centering
    \includegraphics[width=9cm]{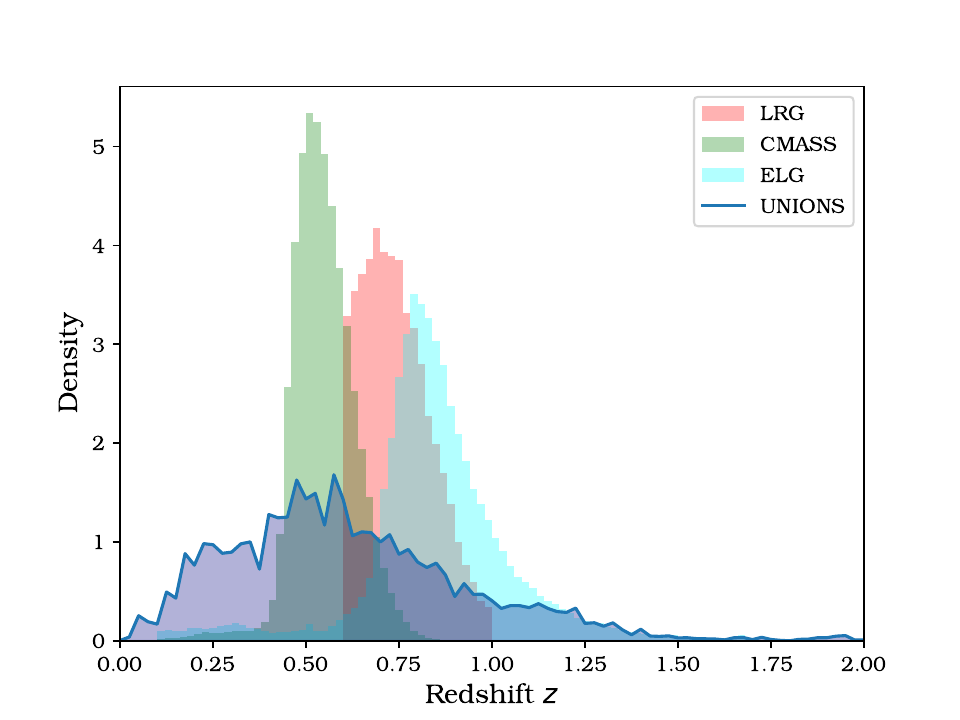}
    \caption{The unweighted normalized redshift distribution $n(z)$ of the three BOSS/eBOSS galaxy samples. The UNIONS $n(z)$ is approximate and for reference only, as blinding was implemented as a shift in the redshift distribution. The $n(z)$ has been obtained by matching galaxies in the CFHTLenS W3 field and calibrated through deep spectroscopic reference samples.}
    \label{redshift}
\end{figure}

In the following, we briefly describe the different BOSS/eBOSS samples used in this paper. The redshift distribution of the different samples is shown in Fig.~\ref{redshift}.

\subsubsection{CMASS}\label{CMASS}

The eBOSS CMASS sample \citep{dawson_baryon_2013} was selected on stellar masses as a passively evolving galaxy sample. It contains galaxies in the redshift range $z\in[0.4,0.8]$. The targets were selected using colour cuts for which the philosophy and functioning are presented in \citet{2001AJ....122.2267E}. These cuts are $i<19.9, d_{\perp} > 0.55$ and $i < 19.86 + 1.6 \times (d_{\perp} - 0.8)$. The quantity $d_{\perp} = (r-i) - (g-r)/8$ quantifies a line in color space perpendicular to the position of the locus of the desired galaxy population at $z > 0.4$. For this sample, the combined weight $w$ is
\begin{equation}
w = (w_\textrm{cp} + w_\textrm{noz} - 1) w_\textrm{sys,tot} w_\textrm{FKP} .
\end{equation}
\subsubsection{LRG}\label{LRG}
The eBOSS luminous red galaxy (LRG) DR16 sample \citep{bautista_sdss-iv_2018} was selected  with the conditions $r - i > 0.98$, and $r - \textrm{W1} > 2.0 \, (r - i)$, where W1 is the $3.4 \mu$m band from the \textit{WISE} survey, and the $r$ and $i$ bands are from the SDSS imaging survey. This results in a sample of high-redshift red galaxies with $z\in[0.6,1.1]$.

The total weight for eBOSS LRGs is \citep{bautista_completed_2020}
\begin{equation}
w = w_\textrm{cp} w_\textrm{noz} w_\textrm{sys,tot} w_\textrm{FKP} .
\label{eq:w_LRG}
\end{equation}

\begin{figure*}\label{maps}
\includegraphics[width=19cm]{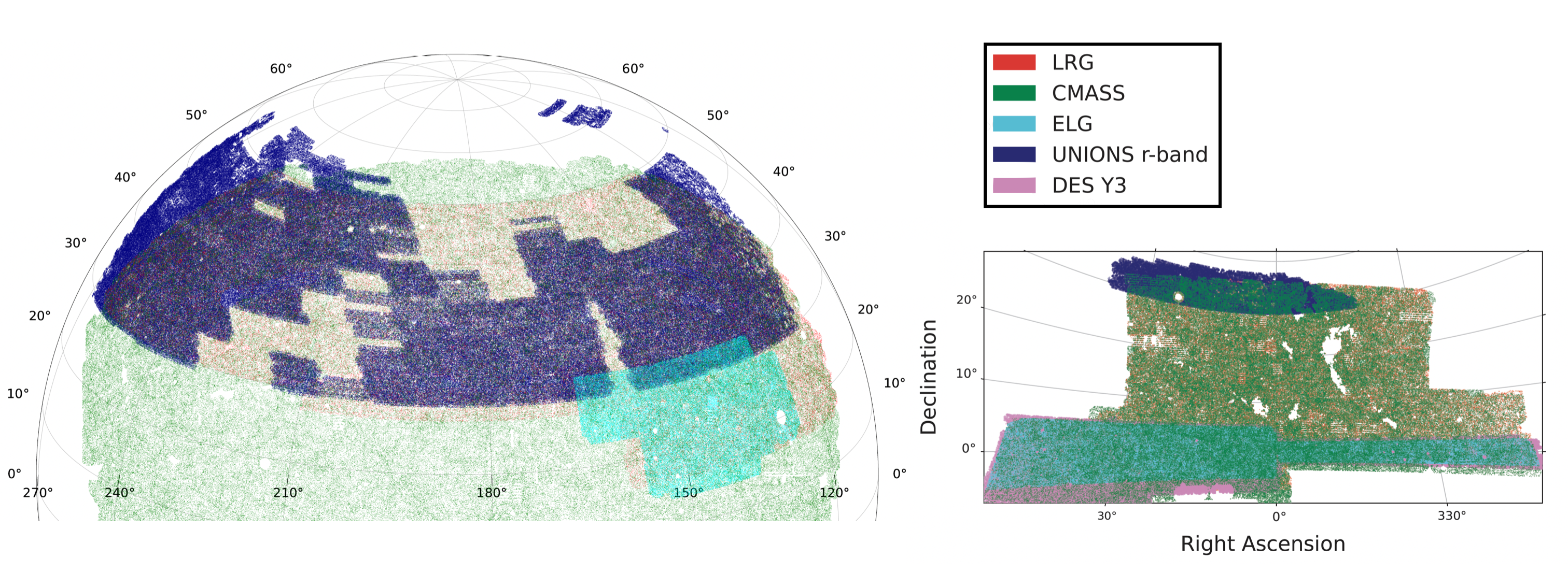}
\caption{Distribution of the different spectroscopic and lensing samples used in this work. The plot on the left shows the Northern Galactic Cap (NGC), where most UNIONS galaxies lie. On the right is a portion of the Southern Galactic Cap (SGC), with the DES Y3 observations in the most southern part. There is no overlap between the UNIONS and DES Y3 catalogues.}
\end{figure*}

\subsubsection{ELGs}\label{ELG}

The eBOSS emission-line galaxy (ELG) sample \citep{raichoor_completed_2020} selected emission line galaxies in the range $0.6<z<1.1$ over an observed sky area of $1,170$ deg$^2$.
The target selection was made via a cut in the $g$ band magnitude, which is sensitive to [$O_{II}$] emitters, and a box selection in the $grz$ color space. In the NGC, the $g$ cut is $21.825 < g < 22.9$. These galaxies form a younger and bluer sample, which means that no intrinsic alignment is expected in these galaxies and that they have a lower clustering bias. The ELG galaxies follows the weight scheme Eq.~\eqref{eq:w_LRG} adopted in eBOSS for the LRG sample.

\subsection{Matching procedure}
\label{sec:matching}

To obtain a sample with precise redshift information, we match the publicly available SDSS catalogues described above with the UNIONS shape catalogue introduced in Sect.~\ref{UNIONS}. We use the $k$-D tree implementation in {\fontfamily{lmtt}\selectfont astropy} with a tolerance of $1\arcsec$, which is the maximum distance between galaxy pairs to be considered a match. We obtain as our shape sample the matched sub-sample of our respective catalogs with precise ellipticity and redshift measurements. For the density sample, we use all galaxies of the respective SDSS catalogs. The different properties of the surveys are described in Table \ref{survey_table}.

\begin{table*}
         
\caption{Sample properties used in this work. The effective redshift $z_{\mathrm{eff}}$ is used for the covariance matrix estimation and described in Sect.~\ref{sec:cov}. The extinction was obtained from the \citet{1998ApJ...500..525S} maps. $\sigma_{\varepsilon,i}$ is the single-component calibrated shape noise, which is also employed in the estimation of the analytical covariance matrix. The fact that the shear responses vary greatly between DES and UNIONS (approx 20\%), but the calibrated $\sigma_{\varepsilon,i}$ are compatible for each population indicates the reliability of the calibration procedure.}    

\centering      
\renewcommand{\arraystretch}{1.3} 

\begin{tabular}{|c | c c c |c c c |}     
\hline\hline       
 & CMASS-UNIONS& LRG-UNIONS &ELG-UNIONS  & CMASS-DES Y3& LRG-DES Y3 &ELG-DES Y3 \\ 
\hline                    
   $N_{\mathrm{gal,shape}}$ & $201,639$ & $78,134$ & $14,762$ & $49,515$ & $22,253$ & $87,929$ \\  
   $N_{\mathrm{gal,dens}}$  & $849,637$ & $174,816$  & $83,769$ & $230,831$ & $67,316$ & $89,967$ \\
   $z_{\mathrm{eff}} $& $0.53$ & $0.74$ & $0.85$ & $0.52$ & $0.75$ & $0.84$ \\
   $\sigma_{e,i}$  & $0.16$ & $0.18$ & $0.17$ & $0.17$ & $0.19$ & $0.17$ \\
   $\bar{n}_{\mathrm{shape}}(z_{\mathrm{eff}}) \ [\mathrm{Mpc^{-3}}]$ & $2.6$ & $0.59$ & $2.2$ & $3.1$ & $0.62$ & $2.4$ \\
   $\bar{n}_{\mathrm{dens}}(z_{\mathrm{eff}}) \ [\mathrm{Mpc^{-3}}]$ & $2.9$ & $0.66$ & $2.4$ & $2.7$ & $0.68$ & $2.5$ \\
   $\langle R \rangle$ & $0.79$ & $0.77$ & $0.91$ & $0.66$ & $0.66$ & $0.78$ \\
   $A_{\mathrm{shape}}$ [deg$^2$] & $2,845$ & $2,589$ & $102$ & $819$ & $682$ & $615$ \\
   $A_{\mathrm{dens}}$ [$\mathrm{deg}^2]$
   & $10,298$ & $5,254$ & $568$ & $2,816$ & $2,099$ & $622$ \\
   $\log\left( \left\langle L \right\rangle / L_0 \right) $ & $-0.056$ & $0.024$ & N.A. & $-0.11$ & $0.006$ & N.A.\\
   Extinction & $0.04$ & $0.04$ & $0.06$ & $0.09$ & $0.10$ &  $0.11$ \\
\hline                  
\end{tabular}
\label{survey_table}
\end{table*}

\section{Correlation and covariance estimations}\label{sec:corr}

\subsection{Shape-density correlation estimator}

To estimate the shape-density correlations of intrinsic galaxy alignment, we use the common projected two-point cross-correlation function $w_{g,+}$ \citep{mandelbaum_detection_2006}.
The intrinsic-alignment measurement is similar to galaxy-galaxy lensing but differs from the latter as the shape tracer can be either at lower or higher redshift compared to the density tracer, at very limited separation.

Following \citet{mandelbaum_detection_2006}, we first compute the 3D correlation function using the modified Landy-Szalay estimator:
\begin{align}
    \xi_\textrm{g+}(r_\textrm{t} , \Pi)
    = \frac{1}{N_{\textrm{r}, \textrm{r}_\textrm{s}}
    (r_\textrm{t}, \Pi)} \times
    \left( \sum_{sd}\gamma_{+,sd} \,
    \Delta_{r_\textrm, \Pi}(\vec x_s - \vec x_d) -  
    \right.
    \nonumber \\
    \left.
    \sum_{sr} \gamma_{+,sr} \,
    \Delta_{r_\textrm{t}, \Pi}(\vec x_s - \vec x_r) \right) .
    \label{eq:xigp}
\end{align}
The variables $s, d$, and $r$ indicate indices of objects in the shape, density, and random sample, respectively. The estimator is a sum of the tangential shear $\gamma_{+, sd}$ 
of objects in the shape sample around the density sample,
corrected by a contribution $\gamma_{+, sr}$ around random points.
The subtraction of the random correlations is introduced to mitigate geometrical effects, for example, shape correlation with the survey mask.

The bin indicator function $\Delta$ of the 2D separation vector
$\vec x = \left( x_\perp, x_\parallel \right)$ is defined as
\begin{equation}
    \Delta_{r_\textrm, \Pi}(\vec x ) = \left\{
    \begin{array}{ll}
        1 & \mbox{if} \;
            \left| \log \left( x_\perp / r_\textrm{t} \right) \right| <
            \frac{\Delta \log r}{2} \quad \mbox{and}
            \left| x_\parallel - \Pi \right| < \frac{\Delta \Pi}{2} \\
        0 & \mbox{else}
    \end{array}
\right. .
\end{equation} 
With this characteristic function, we separate galaxy pairs into logarithmic perpendicular bins with width $\Delta \log r$, and linear radial (line-of-sight) bins with width $\Delta \Pi$.

Next, we integrate the correlation function Eq.~\eqref{eq:xigp} along the line of sight to project out redshift space distortion (RSD) effects stemming from peculiar velocities,
\begin{equation} \label{equ:wgp}
    w_\textrm{g+}(r_\textrm{t}) = \int_{-\Pi_{max}}^{\Pi_{max}}
    \xi_{g+}(r_\textrm{t}, \Pi) \, \textrm{d} \Pi .
\end{equation}
We varied the comoving limit and found a stable value of $\Pi_\textrm{max} = 15$ Mpc, which is consistent with previous works.
In practise, we use 
the {\fontfamily{lmtt}\selectfont Treecorr} package described in \citet{2015ascl.soft08007J} to compute Eq.~\eqref{eq:xigp} in slices of $\Delta \Pi$, and sum up the correlations to obtain $w_{g+}$.


\subsection{Density-density correlation estimator}
We need to estimate a relationship between the total matter power spectrum $P_{\delta}$ and the galaxy power spectrum $P_\textrm{gg}$ as detailed in Sect.~\ref{sec:cluster}. 
To estimate the galaxy bias parameters, we compute the two-point galaxy correlation function using the Landy-Szalay estimator
\begin{equation}
\xi_\textrm{gg}(r_\textrm{t}, \Pi) = \frac{N_{dd}(r_\textrm{t} , \Pi) - 2 N_{dr}(r_\textrm{t} , \Pi) + N_{rr} (r_\textrm{t} , \Pi)}{N_{r,r}(r_\textrm{t} , \Pi)}
\end{equation}

As for the shape-density estimator, the random correlations account for systematic geometrical effects.
We integrate along the line of sight to wash out RSD effects:
\begin{equation} \label{equ:wgg}
    w_\textrm{gg}(r_\textrm{t}) =
    \int_{-\Pi_\textrm{max}}^{\Pi_\textrm{max}} \xi_\textrm{gg}(r_\textrm{t}, \Pi) \, \textrm{d} \Pi .
\end{equation}

\subsection{Shape-Shape correlation estimator}

The majority of the intrinsic alignment constraining power is obtained from $w_\textrm{g+}$, as position-shape measurements are less noisy than shape-shape measurements, due to shape-noise. While $w_\textrm{g+}$ is sensitive to gravitational-intrinsic correlations, as explained in Sect.~\ref{sec:NLA}, the projected shape-shape correlation function $w_{++}$ is sensitive to the intrinsic-intrinsic power spectrum. Previous measurements were presented in \Singh, \Sam.

The corresponding estimator is defined as
\begin{equation}
    \xi_{++}(r_\textrm{t}, \Pi) =
    \frac{1}{N_{\textrm{r}_\textrm{s}, \textrm{r}_\textrm{s}}(r_\textrm{t}, \Pi)}
    \sum_{s s^\prime}\xi_{++,ss^\prime}
    \Delta_{r_\textrm{t}, \Pi}
    (\vec x_s - \vec x_{s^\prime}) ,
\end{equation}
which is then integrated along the line of sight.
\begin{equation}\label{equ:wpp}
    w_{++} = \int_{-\Pi_\textrm{max}}^{\Pi_\textrm{max}} \xi_{++}(r_\textrm{t}, \Pi) \textrm{d} \Pi .
\end{equation}
We include this estimator for completeness and for comparison with previous works.

\subsection{Computing luminosities}\label{sec:lum}

When estimating intrinsic luminosities, various observational effects have to be corrected, which modify the observed galaxy flux. For this we compute $k$-correction using \citep{blanton_k-corrections_2007} as implemented in the python package {\tt{k-correct}}. Further, the $e$-correction is a quantity that reflects the evolutionary processes of galaxies, such as star formation, aging stellar populations, and mergers of galaxies, which change the galaxy brightness as the universe evolves. We obtain the $e$-correction from the \texttt{Ezgal} package \citep{mancone_ezgal_2012}, for which we assume as the stellar population synthesis model the commonly used \citet{bruzual_stellar_2003} model. Lastly, we correct for dust extinction with the $E$-corrections. These values are given by the SDSS catalogue and are based on the \citet{1998ApJ...500..525S} maps.   
To compute the luminosities, we use the {\it{DevFlux}} apertures from the SDSS photometric imaging survey; this allows us to get an approximation of the SED necessary for the $e$-correction. 
The absolute magnitude as a function of the observed SDSS magnitude $r$ is computed with the formula
\begin{equation}
    M_r = r - 5 \ [   \ \mathrm{log}_{10}
        \left(D_\textrm{l}(z)\right) -1 ] - (k+e) + E,
\end{equation}
where $D_\textrm{l}$ is the luminosity distance in pc$/h$.
This equation and the correction quantities are well established for red galaxy populations and can not be generically applied as they require a good understanding of galaxy evolution. Luminosity mean values for different samples are shown in Fig.~\ref{survey_table}. We use this information to split our CMASS galaxies in luminosity sub-samples and compare the luminosity scaling to results obtained from the literature in Sect.~\ref{sec:lum_dep}. Our luminosity estimates are in good agreement with those from \Sam, even though we do not expect exact matches as their measurements are based on the DES $r$ band filter, which differs slightly from SDSS. To obtain $\langle L/L_0 \rangle$ we use the standard pivot luminosity, defined to match an absolute magnitude of $M_r=-22$. 
\section{Modeling intrinsic alignment and galaxy clustering}\label{sec:model}

\subsection{Modelling Galaxy clustering}\label{sec:cluster} 

Galaxies are biased tracers of the underlying dark matter field. At the linear level, the relation is simply $\delta_\textrm{g}(x) = b_1 \delta_\mathrm{m}(x)$, which consequently is for the matter power spectrum
\begin{equation}
P_\textrm{gg}=b_1^2 P_{\delta},
\end{equation}
where $P_{\delta}$ represents the linear matter power spectrum. To go beyond the linear order, we expand the galaxy overdensity as function of the matter density and tidal field up to second order as
\begin{equation} \label{equ:dg}
    \delta_\textrm{g}(x) = b_1 \delta_\mathrm{m}(x) +
    \frac{b_{2}}{2} \left( \delta_\mathrm{m}^2(x) -\langle \delta_\mathrm{m}^2 \rangle \right) + \frac{b_{s}}{2} \left(s^2(x) - \langle s^2 \rangle \right) b_{3 \mathrm{n} 1} \psi+\cdots 
\end{equation}
where $s$ is the tidal tensor, introduced in \citet{mcdonald_clustering_2009}, and expressed in Fourier space as
\begin{equation} \label{equ:tidal} 
    s_{ij}(k) = \left( \hat{k}_i \hat{k}_j - \frac{1}{3} \delta_{ij} \right) \delta_\mathrm{m}(k).
\end{equation}
The number of bias parameters that can be constrained by observations depends on the precision of the data, and the angular scales used in the analysis.
In this study, we only vary $b_1$ and $b_2$, which is sufficient to obtain a good fit. We fix the values of the remaining bias parameters to
\begin{equation}
    b_{s^2} = -\frac{4}{7} \left(b_1 - 1 \right) ; \quad 
    b_{3 \mathrm{nl}} = \left(b_1 - 1 \right) .
\end{equation}
The higher-order terms from Eq.~\eqref{equ:dg} can be used to establish an expansion of $P_\textrm{gg}$ \citep{pandey_alleviating_2020}.
We use {\tt{FASTPT}} \citep{mcewen_fast-pt_2016} to computes the various perturbation contributions in the expansion.

\subsection{Intrinsic alignment models}

Ellipticities are commonly decomposed into three components: the shear $\gamma_\textrm{G}$, the intrinsic ellipticity $\gamma_\mathrm{I}$, and a random contribution $\gamma_\textrm{rand}$,
\begin{equation}
\gamma_{\mathrm{tot}}=\gamma_{\mathrm{G}}+\gamma_{\mathrm{I}}+\gamma_{\mathrm{rand}}.
\end{equation}
The first two components do not account for the entire shape noise. \citet{chisari_cosmological_2013} showed that the predicted dispersion of ellipticities using these terms is two orders of magnitude smaller than the observed value. An additional component $\gamma_{\mathrm{rand}}$ was therefore introduced to quantify unaccounted processes. While not strictly speaking random, these processes do not correlate with the large-scale structure and, therefore, form a noise contribution. In the following sections, we present two models to predict $\gamma_{\mathrm{I}}$.

\subsubsection{NLA}\label{sec:NLA}

The \emph{non-linear alignment} (NLA) model \citep{hirata_intrinsic_2004}  quantifies the contribution of intrinsic alignment to the ellipticity. It is based on \citet{catelan_intrinsic_2001}, stating that the mean intrinsic ellipticities of galaxies forming in a tidal field is related to the Newtonian gravitational potential $\Psi_\textrm{P}$ via the relation
\begin{equation}
\gamma_\textrm{I} = -\frac{C_1}{4\pi G}\left(\nabla_x^2-\nabla_y^2,2\nabla_x \nabla_y \right)\Psi_\textrm{P},
\end{equation}
where the derivatives along $x$ and $y$ are taken in the transverse plane. The constant $C_1$ takes the fiducial value of $5\times 10^{-14} M_\odot\ h^{-2} \mathrm{Mpc}^{-3}$ based on the SuperCOSMOS survey from \citet{brown_measurement_2002}. This ties the tidal deformations to cosmological quantities as the Newtonian potential can be related to the linear matter overdensities $\delta_\textrm{m}$ via the relation in Fourier space
\begin{equation}\label{equ:poisson}
\Psi_\textrm{p}(k) = -4\pi G \frac{\bar{\rho}}{\bar{D}(z)}k^{-2} \delta_\textrm{m}(k),
\end{equation}
where $\bar{D}(z)$ is the rescaled growth factor $\bar{D}(z) \propto (1+z) D(z)$  and $\bar{\rho}$ is the mean density of the Universe.

These relations motivated the expressions of the intrinsic-intrinsic and intrinsic-matter power spectra as 
\begin{equation}\label{equ:Pks}
    P_\textrm{II}(k,z)=F^2(z) P_\delta(k,z) \ ;\ P_{\delta I}(k,z)=F(z) P_\delta(k,z) .
\end{equation}
The function $F(z)$ accounts for the prefactors in the Poisson equation Eq.~\eqref{equ:poisson},
\begin{equation}
    F(z)=-A_{\mathrm{IA}} \, C_1 \, \rho_\textrm{crit} \frac{\Omega_\textrm{m}}{\bar{D}(z)} . 
\end{equation}
Here, $\rho_\textrm{crit}$ is the critical density of the Universe, and $\Omega_\textrm{m}$ the matter density parameter, both expressed at $z=0$.
The $A_{\mathrm{IA}}$ parameter is usually left free to account for a varying amplitude of the intrinsic alignment contribution; in this study, we aim to obtain constraints on this parameter. While initially $P_\delta(k,z)$ in Eq.~\eqref{equ:Pks} was understood as the linear power spectrum, \citet{bridle_dark_2007} instead used the non-linear power spectrum to better account for small-scale processes. While not motivated theoretically, this modification provided a better fit to the data and gave rise to the name \emph{non-linear} alignment model. 

\subsubsection{TATT}\label{sec:TATT}

The \emph{tidal alignment and tidal torque} (TATT) model was introduced as a second-order correction to the NLA model in \citet{blazek_beyond_2019}, analogous to the inclusion of higher-order biases in galaxy clustering.

An expansion in the linear density field and collection of contributing terms allow us to write the unprojected intrinsic ellipticity tensor as
\begin{equation}
    \gamma^\textrm{I}_{ij}(x) = C_1 s_{ij} + C_2 \left( s_{ik}s_{kj} - \frac{1}{3} \delta_{ij} s^2 \right) + C_{1\delta} \left( \delta_\mathrm{m} s_{ij} \right) .
\end{equation}
Summation over repeated indices is implied. 
The most common parametrization established in \citet{blazek_beyond_2019} defines the first parameter as
\begin{equation}
     C_1(z) = -A_1 \bar{C}_1 \rho_{\mathrm{crit}} \Omega_\textrm{m}D^{-1}(z),
\end{equation}
which is the parameter already present in the NLA model. 

The amplitude of the quadratic contribution of the tidal tensor is parameterized by $C_2$, given by
\begin{equation}
    C_{2}(z) = A_2(z) 5\bar{C}_1  \rho_{\mathrm{crit}} \Omega_\textrm{m} {D^{-2}(z)}  .
\end{equation}
The term quantifying the correlation between the tidal tensor and the overdensity is
\begin{equation}
    C_{\delta 1}(z) = b_\textrm{TA} C_1 = -A_{1\delta}(z) \bar{C}_1\rho_{\mathrm{crit}} \Omega_\textrm{m} D^{-1}(z).
\end{equation}
From a theoretical point of view, $b_\textrm{TA}$ can be interpreted as the linear galaxy bias $b_1$. In the TATT model, however, it is an additional free parameter quantifying the correlations between the tidal tensor and the galaxy density field.
Using these parameters one can establish an expression for $P_{\textrm{gI}}$ and $P_{\textrm{II}}$. We refer the reader to \citet{blazek_beyond_2019} for details. 

\subsection{Redshift-space distortions}

Peculiar velocity effects, known as redshift space distortions (RSDs), induce anisotropies along the line of sight, which have to be accounted for in our theoretical model. RSDs are added at the power-spectrum level following the \citet{kaiser_clustering_1987} parametrisation, 
\begin{equation}
P_\textrm{\rm{gg}}^s(\vec k, z) = b_1^2 \left[1 + \beta(z) \mu\right]^2 P_{\textrm{gg}}(k, z),
\end{equation}
with $\mu$ the cosine of the angle between $\vec k$ and the line of sight. The superscript $s$ specifies that the power spectrum is defined in redshift space.
The function $\beta(z) = f(z)/b_1(z)$ is the ratio of 
the growth rate $f(z)$
and the
linear galaxy bias $b_1$.
In the linear regime, only the first three even multipoles are non-vanishing. They are given by \citet{kaiser_clustering_1987} as
\begin{align}
P_0^s(k,z) = & \left(1 + \frac{2}{3} \beta(z) + \frac{1}{5} \beta^2(z) \right) P(k,z); \\
P_2^s(k,z) = & \left( \frac{4}{3} \beta(z) + \frac{4}{7} \beta^2(z) \right) P(k,z); \\
P_4^s(k,z) = & \frac{8}{35} \beta^2(z) P(k,z).
\end{align}
The configuration-space multipoles can then be estimated by the Hankel transform
\begin{equation}
\xi_{\ell}(r) = \frac{i^{\ell}}{2 \pi^2} \int_0^{\infty} k^2 \textrm{j}_\ell(k r) P_{\ell}(k) \mathrm{d} k,
\end{equation}
where $\textrm{j}_\ell$ represent the spherical Bessel functions of order $\ell$.
The 2D anisotropic two-point correlation function can then be reconstructed as 
\begin{equation}
    \xi_\textrm{gg}^s\left(r_{\mathrm{t}}, \Pi\right) =
    \sum_{\ell=0,2,4} \mathcal{P}_{\ell}
    \left(\Pi / d \right)
    \, \xi_{\ell}\left(d\right),
\end{equation}
with $d = \sqrt{r_{\mathrm{p}}^2 +\Pi^2}$ representing the distance between the galaxies.
The estimation of $w_\textrm{gg}$ is then obtained by integrating $\xi_\textrm{gg}^s \left(r_{\mathrm{t}}, \Pi\right)$ along the line of sight. Note that $w_{g+}$ is very mildly affected by RSD's and we do not include these effects here.

\subsection{Predicting the projected correlation functions}

We model the projected correlation functions measured in Eqs.~\eqref{equ:wgp},\eqref{equ:wgg} and \eqref{equ:wpp} via a Limber projection, and using a Hankel transform: 
\begin{align}
\label{eq:w_gp_estimation}
    w_\textrm{gg}\left(r_\mathrm{t}\right)
    =& \int \mathrm{d} z \, \mathcal{W}^{\mathrm{dd}}(z) \int \frac{\mathrm{d} k_{\perp} k_{\perp}}{2 \pi} \textrm{J}_0\left(k_{\perp} r_{\mathrm{t}}\right) P_\textrm{gg}\left(k_{\perp}, z\right); \nonumber \\ 
    w_\textrm{g+}\left(r_{\mathrm{t}}\right)
    =& -\int \mathrm{d} z \mathcal{W}^{\mathrm{ds}}(z) \int \frac{\mathrm{d} k_{\perp} k_{\perp}}{2 \pi} \textrm{J}_2\left(k_{\perp} r_{\mathrm{t}}\right) P_\textrm{gI}\left(k_{\perp}, z\right); \nonumber \\
    w_{++}\left(r_{\mathrm{t}}\right)
    =& \int \mathrm{d} z \mathcal{W}^{\mathrm{ss}}(z) \nonumber \\
    & \times \int \frac{\mathrm{d} k_{\perp} k_{\perp}}{4 \pi}\left\{ \left[ \textrm{J}_0\left(k_{\perp} r_{\mathrm{t}}\right) + \textrm{J}_4\left(k_{\perp} r_{\mathrm{t}}\right)\right] P_{\mathrm{II},\mathrm{EE}}\left(k_{\perp}, z\right) \right. \nonumber \\
    & \left. + \left[ \textrm{J}_0\left(k_{\perp} r_{\mathrm{t}}\right) - \textrm{J}_4\left(k_{\perp} r_{\mathrm{t}}\right) \right] P_{\mathrm{II},\mathrm{BB}}
    \right\}.
\end{align} 
The function $\textrm{J}_\nu$ is the first-kind Bessel function of order $\nu$, and $\mathcal{W}^{ij}$ is the projection kernel \citep{mandelbaum_wigglez_2011} given by:
\begin{equation} \label{equ:kern}
    \mathcal{W}^{ij}(z) = \frac{n^i(z) n^j(z)}{\chi^2(z) \, \mathrm{d} \chi / \mathrm{d} z} \times \left[\int \mathrm{d} z^\prime \frac{n^i(z^\prime) n^j(z^\prime)}{\chi^2(z^\prime) \mathrm{d} \chi / \mathrm{d} z^\prime}\right]^{-1}. 
\end{equation}
$\chi(z)$ is the comoving distance at distance $z$, and $n^{i}(z)$ is the redshift distribution of the shape ($i=$s) or density ($i=$d) sample.

\subsection{Modeling lensing contaminations}

By construction, the estimator $w_\textrm{g+}$ defined in Eq.~\eqref{wgp} is able to pick out intrinsic-alignment correlations at high signal-to-noise. It is, however, also sensitive to a number of other parameters and effects. Some of these effects include a marginalization over cosmological parameters, baryonic feedback, weak gravitational lensing magnification and shape correlations. In this section, we focus on the latter contamination.

To account for the lensing contribution in our measurement, we follow the method of
\citet{joachimi_constraints_2011, tonegawa_first_2022}. 
The galaxy-galaxy lensing contribution is given by the angular power spectrum \citep{joachimi_constraints_2011}
\begin{align}
    C_\textrm{g,G}(\ell; z_1, z_2) = b_1
    \int_0^{\chi_{\textrm{lim}}} \textrm{d} \chi^\prime & \frac{p_n(\chi^\prime|\chi(z_1))
    q_\varepsilon(\chi^\prime| \chi(z_2))}{\chi^{\prime 2}}  
    \nonumber \\
    & \times
    P_\delta \left(\frac{\ell}{\chi'}, \chi' \right) .
\label{equ:cgg}
\end{align}
This requires knowledge of the galaxy bias. The density tracer $p_x(\chi,\chi_1)$ and the lensing kernel $q_\varepsilon(\chi',\chi(z_2))$ are given in full generality for a photometric survey. With spectroscopic surveys, the conditional density tracer $p_x(\chi|\chi_1)$ simplifies to the $n(\chi_1)$ as spectroscopic redshift estimates can be considered exact at the precision we work at. 
This implies that the lensing kernel is expressed as:
\begin{equation}
    q_x(\chi,\chi_1)=\frac{3 H_0^2 \Omega_m}{2c^2}\frac{\chi}{a(\chi)} \int_{\chi}^{\chi_{hor}}d\chi' \ p_x(\chi'|\chi_1) \ \frac{\chi'-\chi}{\chi'}, 
\end{equation}
It can be further simplified by the insertion of a Dirac delta as $p_x(\chi'|\chi_1)$:
\begin{equation}
    q(\chi,\chi_1)=\frac{3 H_0^2 \Omega_m}{2c^2}\frac{\chi}{a(\chi)} \frac{\chi_1-\chi}{\chi_1}
\end{equation}
This allows to rewrite Eq.\eqref{equ:cgg} in the simpler form:

\begin{equation}
    C_{g,G}(\ell,|\chi_1,\chi_2)=b_g  \frac{3 H_0^2 \Omega_m}{2c^2}\frac{n(\chi_1) }{a(\chi_1)} \frac{\chi_2-\chi_1}{\chi_1 \chi_2}  P_\delta \left(\frac{\ell}{\chi_1},\chi_1\right)
\end{equation}
Now we do not express our correlation function in terms of {$\chi_1,\chi_2$}  but use the coordinate system {$r_t,\Pi$} in physical space. To transform the coordinates, we need to start by defining new coordinates which relate the two systems: 
\begin{equation}
    z_m=\frac{1}{2}(z_2+z_1)  \hspace{0.3cm}  ;  \hspace{0.3cm}  r_t=\theta \xi (z_m) \hspace{0.3cm} ; \hspace{0.3cm}  \Pi=\frac{c}{H(z_m)}(z_2-z_1) 
\end{equation}
As we need the correlation functions for a variety of {$z_m,\Pi,r_t$} we build a grid of $\xi(\theta,z_1,z_2)$ in the {$z_1,z_2$} space. We therefore have a grid of correlation functions that we can then use to evaluate the integrated correlation function:
\begin{equation}
    w_{gG}=\int_{0}^{\Pi_{max}}d\Pi \int dz_m W(z_m) \, \xi_{gG}(r_t,\Pi,z_m)
\end{equation}
The kernel here is described in Eq.~\eqref{equ:kern}.
The weak-lensing contributions are of the order of a few percent for each of our samples and are included at the modeling stage. This is consistent with the findings in 
\Sam.

Regarding the other contaminations to the integrated correlation function listed above are sub-dominant on the scales of interest, and we choose, therefore, not to model them here. The marginalization over cosmological parameters has been studied and we refer the reader to \citet{johnston_kidsgama_2019}.

\subsection{Covariance matrix}\label{sec:cov}
\subsubsection{Analytical covariance matrix}

To estimate the covariance matrix for our different estimators, we use an analytical model, which was first described in \citet{chisari_cosmological_2013}, and used in \citet{samuroff_advances_2021} and \Sam.
To compute the covariance matrix we follow the procedure laid out in \cite{samuroff_advances_2021}.
We compute the full covariance matrix of the joint vector containing the integrated correlation functions $w_\textrm{g+}, w_\textrm{gg}, w_{++}$. Each block of the analytical covariance matrix can be decomposed into three types of contributions: a Gaussian part that accounts for cosmic variance, a shot noise part, and a mixed term.

The full general equation for the covariance matrix between two integrated correlation function estimates $w_{\alpha\beta}$ and $w_{\gamma\varepsilon}$ for $\alpha, \beta, \gamma, \varepsilon \in [\textrm{g}, +]$ at angular scales $r_{\textrm{t}, i}$ and $r_{\textrm{t}, j}$ is given by
\begin{multline}\label{equ:cov}
\operatorname{Cov}\left[w_{\alpha \beta}\left(r_{\mathrm{t}, i}\right) w_{\gamma \varepsilon}\left(r_{\mathrm{t}, j}\right)\right] = 
 \frac{1}{\mathcal{A}\left(z_\mathrm{eff}\right)} \int_{0}^{\infty} \frac{k \mathrm{~d} k}{2 \pi} \bar{\Theta}_{\alpha \beta}\left(k r_{\mathrm{t}, i}\right) \bar{\Theta}_{\gamma \varepsilon} \left(k r_{\mathrm{t}, j}\right) \\ 
\times \Big[\Big.(P_{\alpha  \gamma}(k) + \delta_{\alpha  \gamma} N^{\alpha}) \times (P_{\beta \varepsilon}(k) + \delta_{\beta \varepsilon} N^{\beta})\\ 
+ (P_{\alpha \varepsilon}(k) + \delta_{\alpha \varepsilon} N^{\alpha}) \times (P_{\beta \gamma}(k) + \delta_{\beta \gamma} N^{\beta}) \Big. \Big].
\end{multline}
The prefactor is the inverse of $\mathcal{A}\left(z_{\mathrm{eff}}\right)$, the comoving area at redshift $z_{\mathrm{eff}}$. To obtain this effective redshift, we take the average of the sample redshift distribution $n(z)$, weighted by the square of the number of galaxy pairs  $N^2_{\mathrm{gal}}(z)$. This is motivated by the fact that the integrated correlation function $w_{\alpha\beta}$ is sensitive to the number of galaxy pairs. When the shape and density samples differ for $w_{g+}$, we use the effective area of the smaller sample, as this is where pairs are formed. The effective redshift approximation is accurate for our samples but might need to be refined for more complex redshift distributions. We will improve upon this approximation in future work.

The $\bar \Theta$ functions are bin-averaged Bessel functions,
\begin{equation}
\bar{\Theta}_{\alpha \beta}\left(k, r_{\mathrm{t}, i}\right) = \int_{r_{\mathrm{t}, i}^{\mathrm{min}}}^{r_{\mathrm{t}, i}^{\mathrm{max}}}\ 2 \pi r\, \Theta_{\alpha \beta}\left(k, r \right) \textrm{d} r .
\end{equation}
with
$\Theta_\textrm{gg} = \textrm{J}_0$,
$\Theta_\textrm{g+} = \textrm{J}_2$
and
$\Theta_\textrm{++} = \textrm{J}_0 + \textrm{J}_4$
.

The two different shot-noise terms are $N^\textrm{g}= 1 / \bar{n}_{\mathrm{density}}(z_\mathrm{eff})$ corresponding to Poisson noise from discrete galaxy positions, and $N^{+} = \sigma_\varepsilon^2/\bar{n}_{\mathrm{shape}}(z_\mathrm{eff})$ from intrinsic galaxy shapes\footnote{Note that \Sam \ indicates $1/N_{\textrm{pairs}}$ as their shot noise, which we found to be a negligible contribution.}. We use $\sigma_\varepsilon$ as the single-component shape noise term defined in \citet{gatti_dark_2021} as 
\begin{equation}
\sigma_e^2 = \frac{1}{2} \left[\frac{\sum_i \left(w_i e_{i, 1}\right)^2}{\left(\sum_i w_i\right)^2} + \frac{\sum_i \left(w_i e_{i, 2}\right)^2}{\left(\sum_i w_i\right)^2}\right] \left[\frac{\left(\sum_i w_i\right)^2}{\sum_i w_i^2}\right] .
\end{equation}

The estimator with the highest signal-to-noise ratio carrying the bulk of information on intrinsic-alignment parameters is the shape-density integrated correlation function $w_\textrm{g+}$ defined in Eq.~\eqref{eq:w_gp_estimation}. We explicitly write out the covariance Eq.~\eqref{equ:cov} for the special case of $(\alpha\beta) = (\gamma\varepsilon) = (\textrm{g}+)$ as
\begin{multline} \label{eq:cov_wgp}
\operatorname{Cov}\left[w_{g+}\left(r_{\mathrm{t}, i}\right) w_{g+}\left(r_{\mathrm{t}, j}\right)\right]= 
 \frac{1}{\mathcal{A}\left(z_\mathrm{eff}\right)} \int_{0}^{\infty} \frac{k \, \mathrm{d} k}{2 \pi} \bar{\Theta}_{\alpha \beta}\left(k r_{\mathrm{t}, i}\right) \bar{\Theta}_{\gamma \varepsilon} \left(k r_{\mathrm{t}, j}\right) \\
\times \left[\left(P_\textrm{gg}(k) + \frac{1}{\bar{n}_{\mathrm{density}}(z_\mathrm{eff})}\right) \times \left(P_\textrm{II}(k) + \frac{\sigma_\varepsilon^2}{\bar{n}_{\mathrm{shape}}(z_\mathrm{eff})}\right) 
+ P^2_\textrm{gI}(k)  \right].
\end{multline}
The mixed power spectrum $P^2_\textrm{gI}(k)$ does not have a shot-noise contribution as it corresponds to a cross-correlation, resulting in the corresponding Kroenecker deltas in Eq.~\eqref{equ:cov} to vanish.

We plot the diagonal of this covariance in Fig.~\ref{fig:cov_terms}.
On small scales, the covariance matrix is shot-noise dominated. On larger scales, the mixed term of the clustering power spectrum and galaxy shape noise becomes the leading term. Note that this mixed term is missing from the covariance matrix of \Sam\ (their Eqs.~26 and 27). Since this term is dominant for $r_\textrm{t} \gtrapprox 20$ Mpc, this omission resulted in a bias of their results, and an underestimation of the error bars published in \Sam. We confirmed this finding by applying our analytical covariance matrix without this mixed term to DES Y3 data, leading to the same biases as in \Sam.
Note that this term was still present in the earlier publication \cite{samuroff_advances_2021}

In Sect.~\ref{sec:results} we present a re-analysis of the DES Y3 and BOSS correlations and show the corrected intrinsic-alignment constraints.

\subsubsection{Jackknife covariance matrix}

To validate our analytical covariance matrix, we compare it to a data resampling covariance using the jackknife method (\Singh, \citealt{johnston_kidsgama_2019},  \citealt{fortuna_kids-1000_2021}, \citealt{joachimi_constraints_2011}). We split the observed area into $n_\textrm{s}$ patches using the $K$-means algorithm implemented in $\tt{Treecorr}$. A number of $n_\textrm{s}$ realisations are built by removing one patch from the data set at a time. The correlation function is computed on the remaining data. The idea is that the patches contain the largest scales probed by the correlation function. By removing a patch, we can estimate the variation in the data without any external analytical formulation or simulation. The jackknife estimation of the covariance matrix is given by:
\begin{equation} 
    \hat{C}_{ij} =\frac{n_{\mathrm{s}}-1}{n_{\mathrm{s}}} \sum_{k=1}^{n_{\mathrm{s}}}\left(x_{i}^{k}-\bar{x_{i}}\right)\left(x_{j}^{k}-\bar{x_{j}}\right)
\end{equation}
with $x_{i}^{k}$ the $k$-th correlation function computed from the $n_\mathrm{s}$ jackknife samples. The $ij$ indices are the bins in $r_\mathrm{t}$ between which we want to compute the covariance. The average $\bar{x}$ is taken over the $n_\textrm{s}$  jackknife estimations. 

\begin{figure}
    \centering
    \includegraphics[width=1.\linewidth]{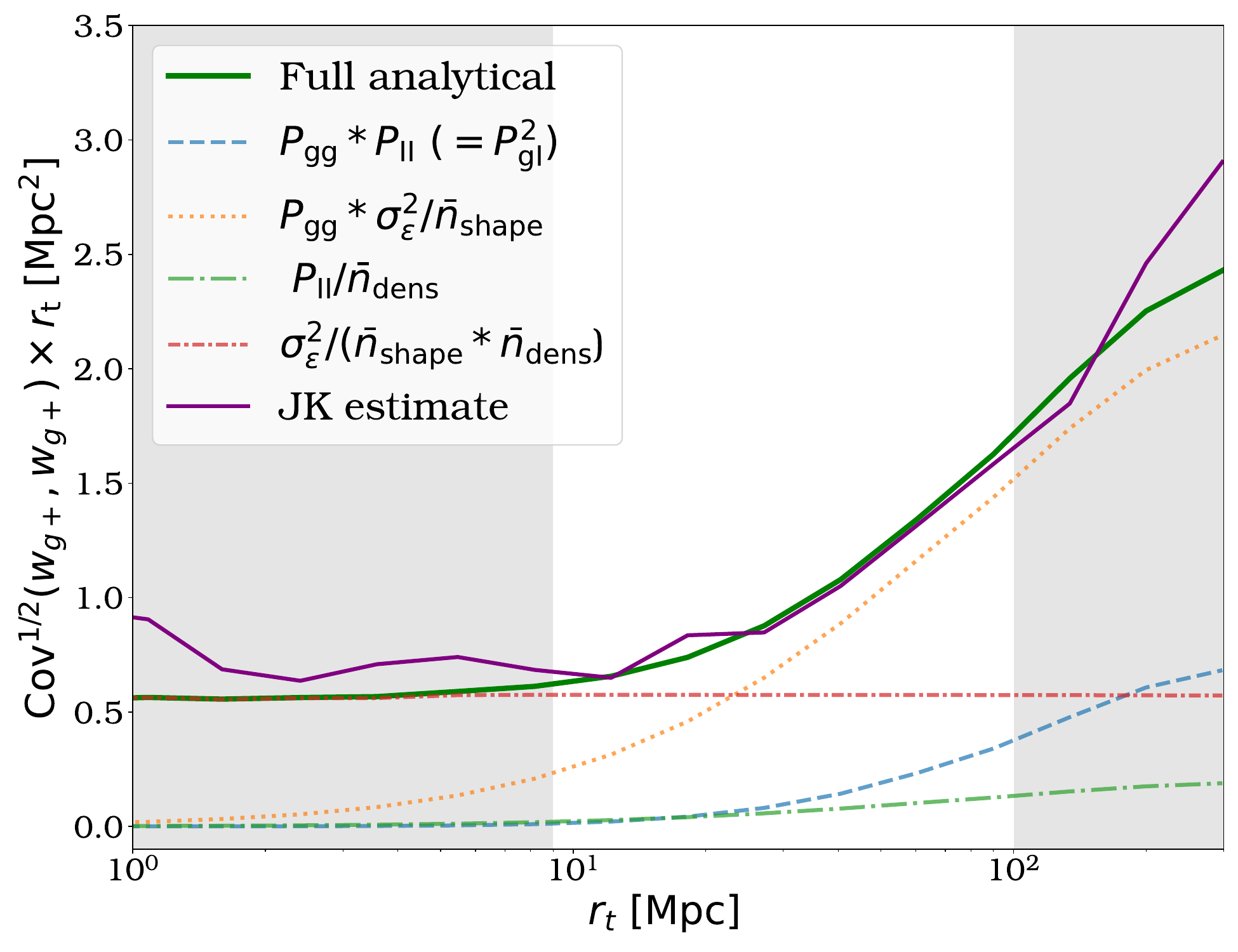}
    \caption{
    The square root of the diagonal of the analytical covariance matrix of $w_{\mathrm{g}+}$ for the CMASS-UNIONS sample
     (green solid line). The individual contributions are
    cosmic variance (dashed; the $P_\textrm{gg}P_\textrm{II}$ term is identical to the $P_\textrm{gI}$ contribution); mixed clustering - ellipticity shot noise (dotted); mixed intrinsic - density shot noise (long dash-dotted); and pure shot noise (short dash-dotted). See Eq.~\eqref{eq:cov_wgp} and text for details. For comparison, we plot the jackknife variance (purple solid line), which is in excellent agreement with the analytical covariance over the relevant scales.
    }
    \label{fig:cov_terms}
\end{figure}

Fig.~\ref{fig:cov_terms} shows the excellent agreement between the analytical and jackknife covariance over the angular scales of interest. Since the jackknife covariance is noisy, in particular on the off-diagonal elements \citep{samuroff_advances_2021}, we used the analytical covariance for the parameter inference.

\subsection{Scale cuts}\label{scale_cuts}

As described in Sect.~\ref{BOSS_cat}, the eBOSS samples were originally selected for galaxy clustering on large scales, with the explicit goal of measuring the baryonic acoustic oscillation scale around $100 \, h^{-1}$ Mpc. The sample is, therefore, not optimized for small-scale correlations where intrinsic alignment becomes most relevant. To include smaller scales requires a more densely sampled survey such as GAMA or DESI.

Furthermore, we do not include small scales in our analysis as baryonic feedback and non-linearities can become dominant on these scales. To be consistent with previous works we introduce a small-scale cut at $9$ Mpc. We also exclude points above $100$ Mpc from the fit. This is motivated by the projection effects described in \citet{singh_intrinsic_2016}: The integration over $\Pi$ along the line of sight causes a suppression for large $\Pi$, as galaxy shapes pointing towards the density tracer at large line-of-sight separation will appear smaller along the radial axis when projected on the 2D observer plane.
While this projection effect exists at all transverse separations $r_\mathrm{t}$, it is sub-dominant where the amplitude of $w_{\mathrm{g}+}$ is large. The sensitivity of $w_{g+}$ to $P_\textrm{gI}$ in the $[r_T,\Pi]$ plane is quantified in \citet{joachimi_constraints_2011}. To mitigate these projections effect, the 3D Yamamoto estimator \citep{kurita_power_2021} or the real-space 3D correlator \citep{singh_increasing_2024} can be used, which circumvent the issue by using the full 3D information.

For consistency with previous works, we take $21$ log-spaced bins in the range $r_\mathrm{t} \in [0.1, 350 ]\, \mathrm{Mpc}$ to compute our projected correlation functions.

For the cosmological parameters we assumed $h=0.69$, $n_\mathrm{S}=0.97$, $A_\mathrm{S}=2.15\times10^{-9}$ and $\Omega_\mathrm{m}=0.3$.
We use the \texttt{emcee} package to fit our models. All correlation functions are computed with the {\tt{TreeCorr}} package \citep{2015ascl.soft08007J}.

\section{Results} \label{sec:results}
\subsection{Integrated correlation functions}

\begin{figure*}
\hspace*{-0.3cm}
\includegraphics[width=1.\textwidth]{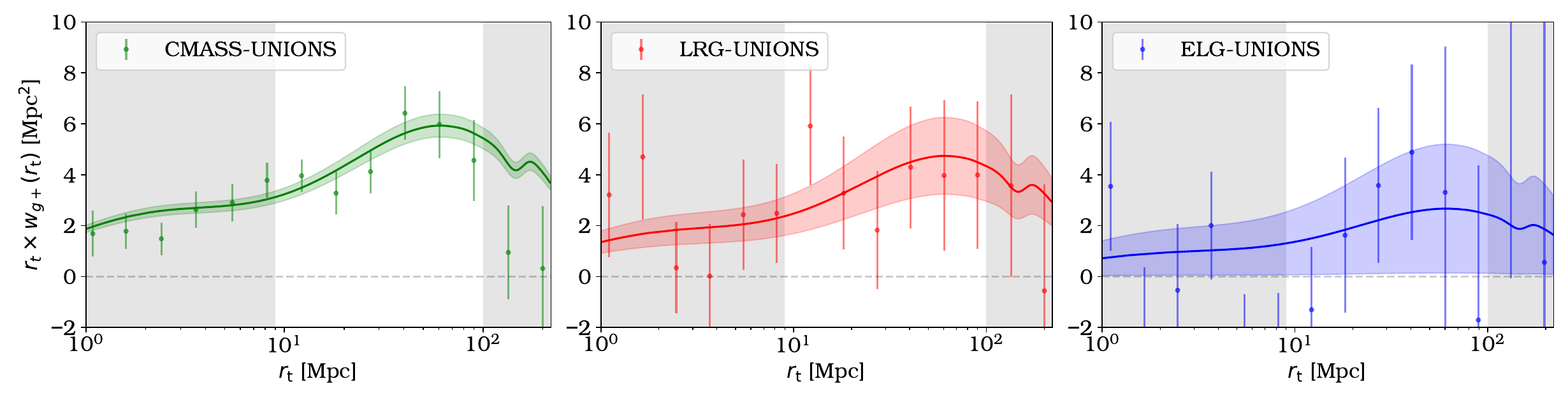} \\
\hspace*{-0.3cm}
\includegraphics[width=1.\textwidth]{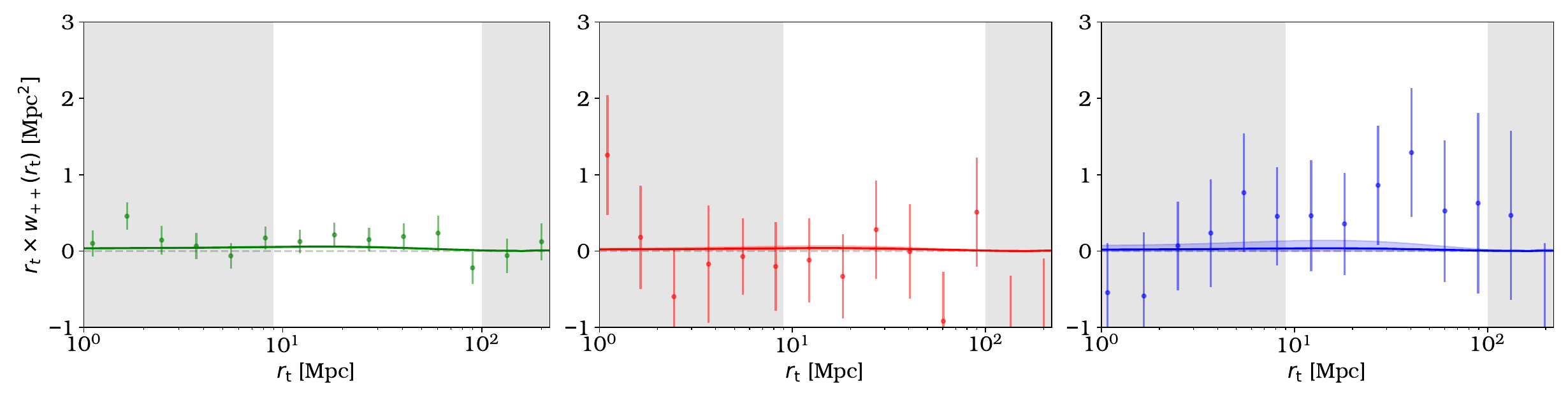}

\caption{The measurements of $w_\textrm{g+}$ \emph{(upper panels)} and $w_{++}$ \emph{(lower row)} for the UNIONS-CMASS \emph{(left)}, UNIONS-LRG \emph{(middle)}, and UNIONS-ELG \emph{(right column)} samples. The solid lines show the best-fit NLA model, where the shaded areas indicate the $1\sigma$ uncertainty associated with the fit. The grey areas are excluded from the fit as explained in Sect.~\ref{scale_cuts}.} 
\label{wgp}
\end{figure*}

We show in Fig.~\ref{wgp} the measurement of $w_\textrm{g+}$ and $w_{++}$ for the different UNIONS-BOSS/eBOSS samples as described in Sect.~\ref{sec:cat}. 
We observe strong intrinsic alignment in our two red galaxy samples, CMASS and LRG, consistent with what has been shown in the literature. In grey, we shade the areas excluded from the fit, as motivated in Sect.~\ref{scale_cuts}. 
For CMASS-UNIONS, our measured $w_\textrm{g+}$ is well fitted by the NLA model down to scales of $r_\textrm{t}\approx 5$ Mpc. Below this scale, the CMASS galaxies are sparsely sampled, and we therefore do not see the appearance of a $1$-halo term. Overall we remain on linear scales where the NLA model performs well. The $w_{++}$ plot shows an apparent positive correlation, which seems under-fitted by the NLA model. The reduced $\chi^2$ summarised in Table \ref{tab:res} indicates that the fit is good, but the null hypothesis is also not excluded for this data vector. Our fit of $A_\textrm{IA}=4.02^{+0.31}_{-0.31}$ represents one of the most significant detections in the literature. 

In the LRG sample, the picture is similar, although the measurement is slightly more scattered. The fit follows the data points well down to scales of $r_{T}\approx 8$ Mpc. The large error bars are due to the scarce density of the LRG galaxies, which in turn are responsible for the large uncertainty on the $A_{\mathrm{IA}}$ fit, even though the 80,000 galaxies form a consequential sample for this type of measurement. The $w_{++}$ correlation function is noisy and has negative points. Its error bars are shot-noise dominated and we do not give any strong interpretation of $w_{++}$  as the measurement is compatible both with the NLA model and with the null hypothesis. The total fit gives $A_\textrm{IA}=3.3^{+1.0}_{-1.0}$, a value we compare to other measurements in Sect.~\ref{sec:compa}. 

For the ELG sample, we expected no intrinsic alignment signal as this is a sample mainly composed of blue galaxies, which have been shown to exhibit no intrinsic alignment. While we measure a relatively high value of $A_\textrm{IA}$, the error bars are still compatible with $A_\textrm{IA}=0$ at the 1.1$\sigma$ level. The $w_{++}$ vector also appears to exhibit a strong correlation on larger scales, but here again, the points are correlated, and the data vector is compatible with the null hypothesis.

All results are summarized in Table \ref{tab:res}. The global reduced $\chi^2$ indicates that our overall fit is good. Note that while we report the individual reduced $\chi^2$ for each data vector separately, they are for reference only as the fit is performed jointly on all 3 data vectors, and the covariance between them is not negligible. We consider 5 degrees of freedom for $w_{++}$ and $w_{\mathrm{g+}}$ (6-1), 6 for $w_{\mathrm{gg}}$ (8-2) and 17 for the total fit (20-3).

\begin{table*}

\caption{Mean and standard deviations of the posterior distributions of the 3 parameters and the reduced $\chi^2$ values associated with the diagonal blocks of the covariance matrix. Degrees of freedom are described in the text. The variation in galaxy bias parameters between UNIONS and DES can be explained by the different footprints used for the density tracers.}

\label{tab:res}

\renewcommand{\arraystretch}{1.3}
\begin{tabular}{|c | c c c |c c c |}     
\hline\hline       
 & CMASS-UNIONS& LRG-UNIONS &ELG-UNIONS  & CMASS-DES Y3& LRG-DES Y3 &ELG-DES Y3 \\ 
\hline                    
   $A_1$ & $4.02^{+0.31}_{-0.31}$ & $3.3^{+1.0}_{-1.0}$ & $3.1^{+3.3}_{-2.9}$  & $3.18^{+0.64}_{-0.64}$  & $6.4^{+2.0}_{-1.7}$ & $-0.6^{+1.0}_{-1.0}$\\  
   $b_1$  & $2.06_{-0.01}^{+0.01} $&$ 2.21^{+0.03}_{-0.03}   $&$ 1.345^{+0.07}_{-0.77}  $&$ 2.00^{+0.02}_{-0.02} $&$ 2.22^{+0.04}_{-0.03} $&$ 1.38^{+0.07}_{-0.05}$\\
   $b_2$  &$ -0.09_{-0.09}^{+0.1} $&$ 0.76^{+0.38}_{-0.28}     $&$ -0.43^{+0.90}_{-0.81}  $&$ 0.19^{+0.19}_{-0.16} $&$ 0.18^{+0.97}_{-0.55} $&$ -0.74^{+0.57}_{-1.1}$\\
   $\chi^2 (w_{gg}) $&$ 2.24 $&$ 0.57    $&$ 0.72$&$ 0.39$&$ 0.53 $&$ 1.26$ \\
    $\chi^2 (w_{g+}) $&$ 0.87 $&$ 0.48    $&$ 0.39 $&$ 1.59$&$ 1.17 $&$ 0.81$\\
    $\chi^2 (w_{++})$&$0.80$&$ 0.31 $&$ 0.34 $&$0.21$&$0.47$&$ 0.53$\\
    Total $\chi^2 $ &$1.40$&$ 0.43 $&$ 0.57 $&$ 0.71 $&$0.67$&$0.84$\\
   
\hline                  
\end{tabular}

\end{table*}

\subsection{Model comparison}
\begin{figure}
\hspace*{-0.8cm}
\includegraphics[width=0.55\textwidth]{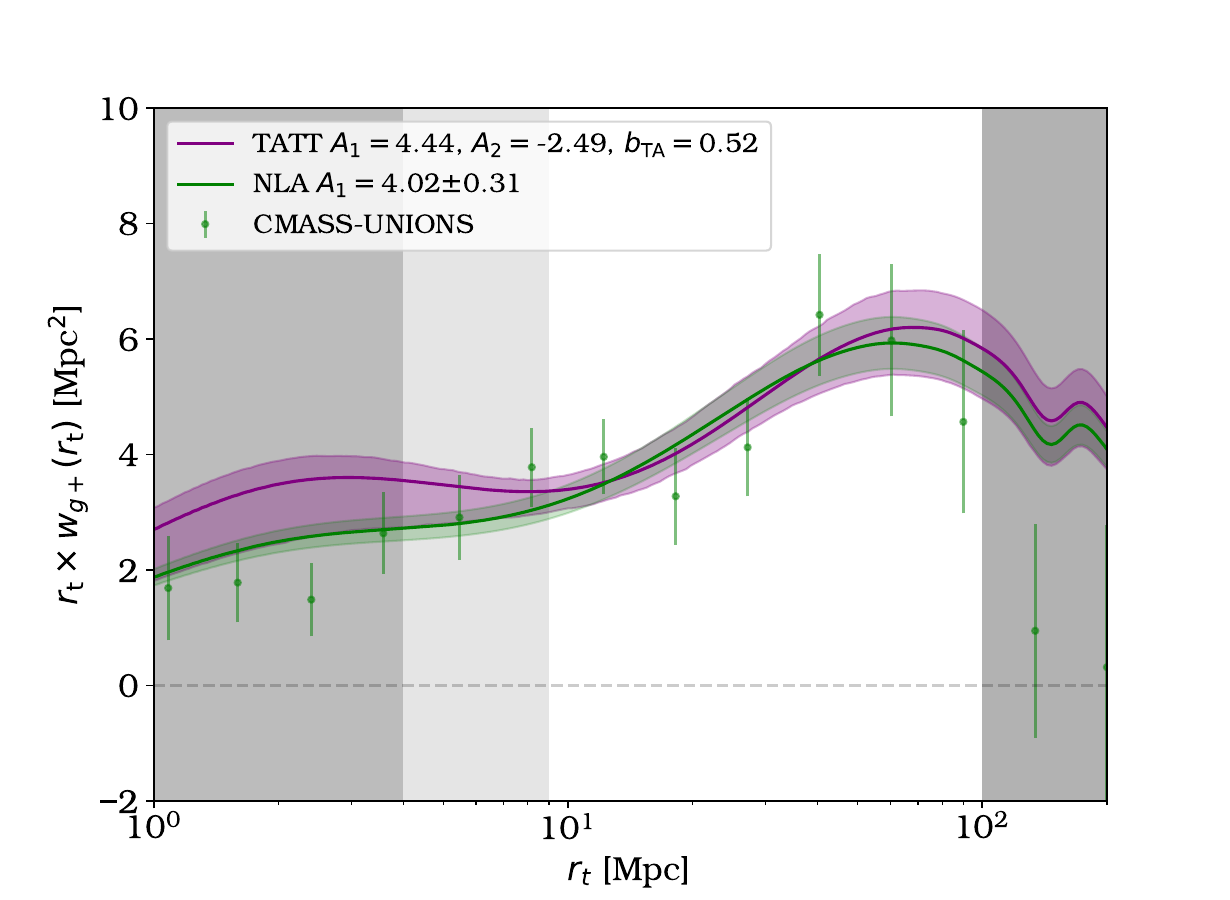}
\caption{The best-fit lines and 1$\sigma$ contours for the NLA and TATT, respectively. The TATT model is fit down to 4 Mpc  but both models appear very consistent. The 1$\sigma$ TATT contours are obtained by drawing 500 models from the posterior and calculating the 68\% confidence interval at eat $r_t$
}
\label{TATT1sig}
\end{figure}

\begin{figure}
\hspace*{-0.8cm}
\includegraphics[width=0.55\textwidth]{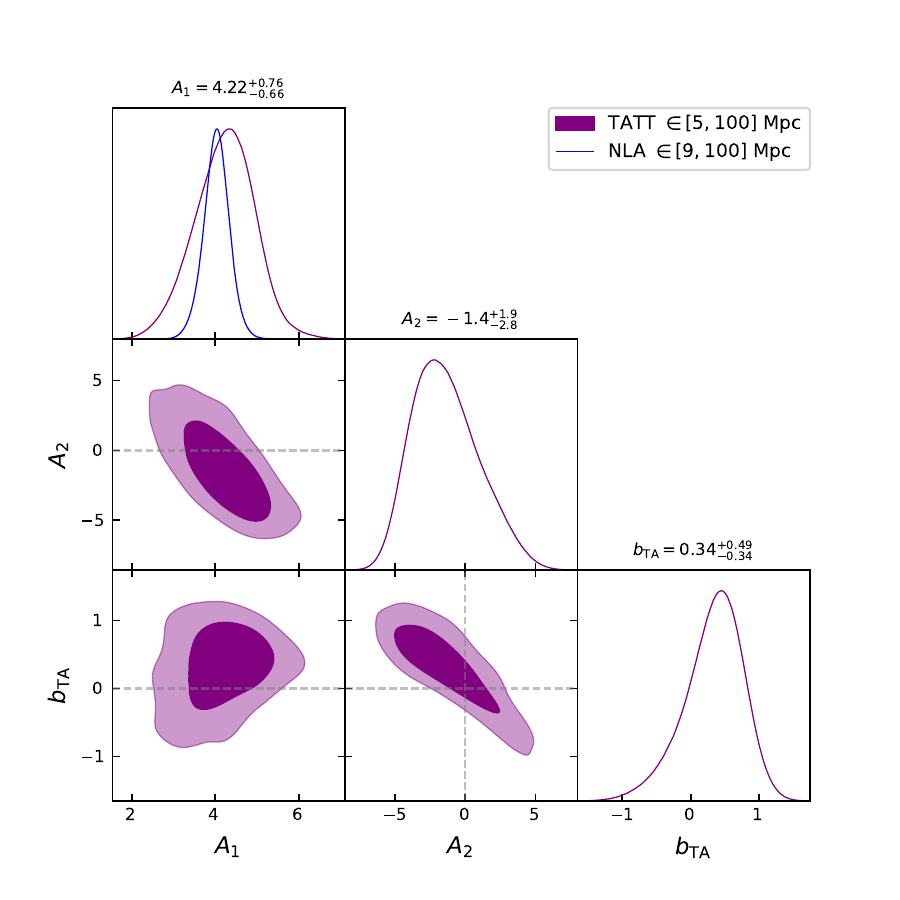}
\caption{Posterior contours of the NLA and TATT model fitted to the CMASS-UNIONS sample. We see no deviation from the NLA case as $A_2$ and $b_{\mathrm{TA}}$ appear consistent with 0.}
\label{fig:TATTmcmc}
\end{figure}

With our high SNR measurement, we test the TATT model \citep{blazek_beyond_2019} to see if the additional degrees of freedom can help in modelling the $w_\textrm{g+}$ signal. The TATT model is a higher-order expansion, as explained in Sect.~\ref{sec:TATT}, and we therefore fit $w_{\mathrm{g}+}$ and $w_{++}$ down to 4 Mpc.
The constraints on the second-order TATT parameters from CMASS-UNIONS can be seen in Fig.~\ref{fig:TATTmcmc}. From Fig.~\ref{TATT1sig}, the NLA and TATT best-fit models are in very good agreement, within $1\sigma$ down to small scales. It appears that the scales outside the fitted range are better captured by the NLA model. Considering the reduced $\chi^2$ we get,
\begin{equation}
    \mathrm{NLA: } \ \chi^2 = 1.40; \\
    \mathrm{TATT: } \ \chi^2 = 1.32 .\\
\end{equation}
These reduced $\chi^2$ have been obtained for the full data vectors meaning, 8+6+6-3 degrees of freedom for NLA and 8+8+8-5 degrees of freedom for TATT. To compare them we can use the bayesian information criterion (hereafter BIC) defined as 
\begin{equation}
\mathrm{BIC}=k ~ \mathrm{log} (N_\mathrm{p}) +\chi^2,
\end{equation}
where $k$ is the number of parameters and $N_\mathrm{p}$ is the number of fitted points. We respectively obtain:
\begin{equation}
    \mathrm{NLA: } \ \mathrm{BIC} = 32.8; \\
    \mathrm{TATT: } \ \mathrm{BIC} = 40.7\\.
\end{equation}
From $\Delta_{\mathrm{BIC}}(\mathrm{TATT}-\mathrm{NLA})=7.9$ we conclude that the additional TATT degrees of freedom are not needed here. We can nevertheless make a few observations:
First, the $A_1$ and $A_2$ parameters are strongly degenerate. This is expected as $A_1$ is the proportionality factor for $s_{ij}$, and $A_2$ for $s_{ij}^2$, displaying similar behaviour for small values of the tidal tensor. To decorrelate the two parameters, the largest possible variations of the tidal field must be probed so that the linearisation breaks down and the quadratic regime can be distinguished from the linear regime. It is not obvious if this will happen on scales relevant for weak lensing, since we are in the perturbation-theory domain where variations are small. 

Further, we observe a degeneracy between $b_\textrm{TA}$ and $A_2$. As $b_{TA}$ is multiplied with the overdensity $\delta_\mathrm{m} s_{ij}$, a similar argumentation holds for these two parameters as for the $(A_1, A_2)$ pair. From theoretical arguments $b_\textrm{TA}$ can be expected to be comparable to $b_1$ as it correlates the tidal tensor with the overdensity field, traced here by galaxies. The fact that we constrain $b_{\mathrm{TA}}$ to values compatible with zero, far away from $b_1\approx2$, can be an indication that the linear assumptions allowing us to take the product of the linearised density field with the tidal tensor can not be naively applied here. 

Finally, we see no correlation between $b_\textrm{TA}$ and $A_1$. This can be interpreted by the model picking up on the density field and decorrelating it from the tidal tensor. As both parameters are multiplied with each other to form the parameter $A_\textrm{d}$, the fact that raising one is not simply equivalent to lowering the other indicates that the measurement effectively decorrelates matter variations from tidal-field modifications. Upcoming measurements will allow us to go down to smaller scales and properly understand the behaviour of these parameters for different samples.

\

\begin{figure*}
    \centering
    \hspace*{-0.55cm}
    \includegraphics[width=1.03\linewidth]{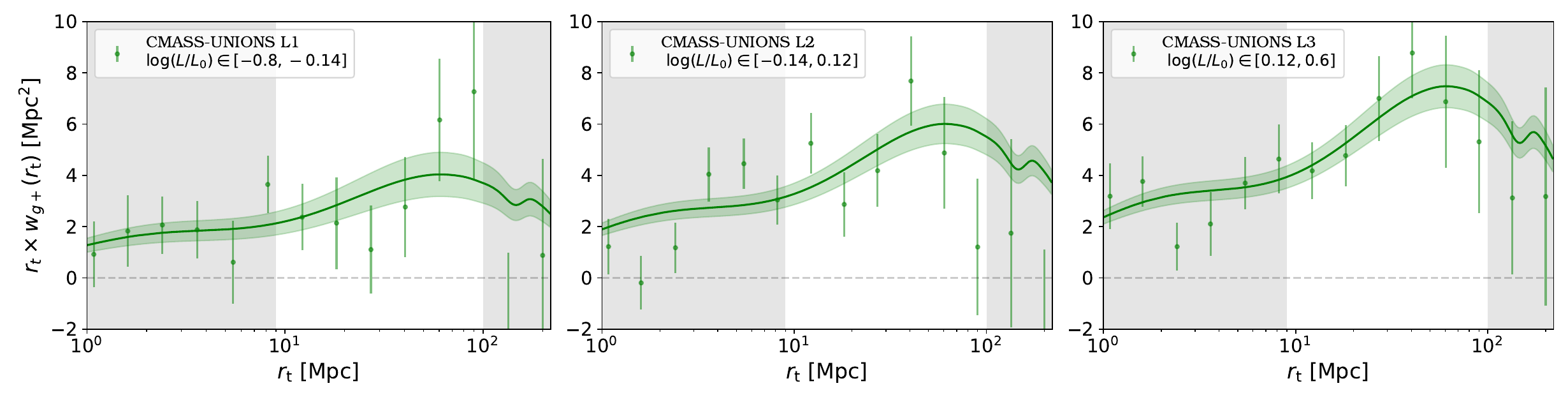}
    \caption{Integrated correlation function $w_\textrm{g+}$ as measured on the CMASS-UNIONS sample split in $3$ bins of luminosity, showing increasing luminosity from left to right panels. One can see a clear increase in the amplitude of the signal.}
    \label{fig:wgp_lum}
\end{figure*}

\subsection{Luminosity Dependence}\label{sec:lum_dep}

We measure the luminosity scaling of our intrinsic-alignment measurements of red galaxies and compare the results to previous studies. As described in Sect.~\ref{sec:lum}, multiple observations have shown a correlation between luminosity and intrinsic alignment. Such a correlation is motivated by recent findings in \citet{fortuna_kids-1000_2024}, who showed that luminosity scales with halo mass. This explains why luminosity is a good predictor of intrinsic alignment, as halo masses determine the amplitude of the tidal field. In Fig.~\ref{fig:wgp_lum} we plot our UNIONS  measurements for the CMASS sample split into three equi-populated bins with limits $\log (L/L_0) \in [-0.8;-0.14;0.12;0.6]$. The intrinsic-alignment amplitude shows a clear increase with luminosity.

\begin{figure}
\hspace*{-1.3cm}
\includegraphics[width=0.6\textwidth]{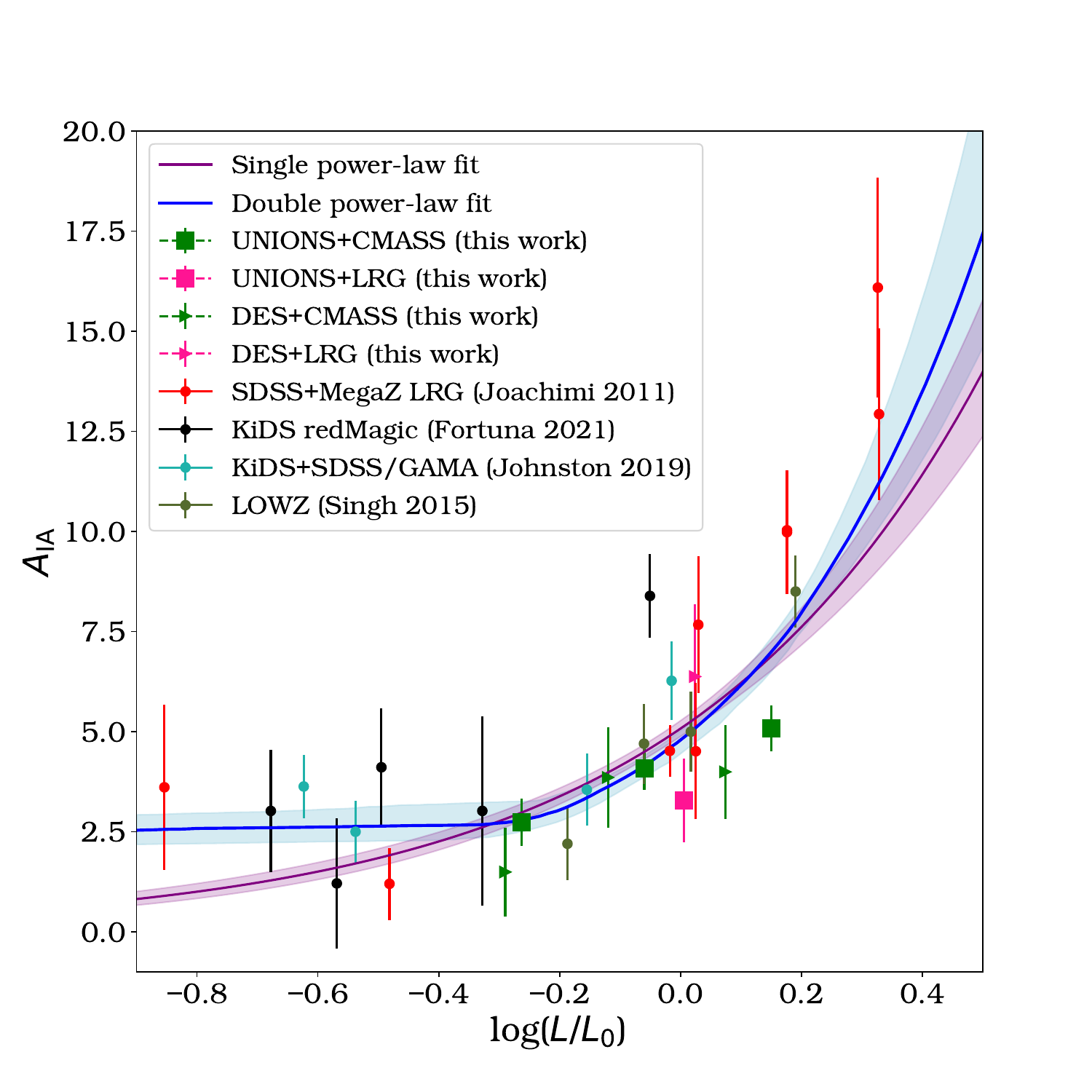}
\caption{Intrinsic alignment amplitude as a function of luminosity as measured by different surveys. The fitted function is $\alpha \langle L/L0\rangle^{\beta}$. We can see very similar scaling with luminosity when comparing our CMASS-DES and CMASS-UNIONS measurements.}
\label{fig:lumi}
\end{figure}

Fig.~\ref{fig:lumi} shows the NLA amplitude $A_\textrm{IA}$ from our UNIONS and DES measurements with the CMASS and LRG samples, together with a collection of data points from the literature.
Our samples are in agreement with measurements from  \citet{joachimi_constraints_2011}; \Singh; \citet{johnston_kidsgama_2019, fortuna_kids-1000_2021}. The CMASS-UNIONS points are slightly higher than the CMASS-DES points, but both seem to scale in the same way with luminosity. On the contrary, the LRG-UNIONS point is lower than the LRG-DES point. We will come back to this discrepancy in Sect.~\ref{sec:compa}.

To quantify the luminosity-dependence and evolution of the intrinsic-alignment amplitude, we fitted all data points from Fig.~\ref{fig:lumi}, assuming their independence. Various functional forms have been tested; \cite{joachimi_constraints_2011} introduced the scaling relation
\begin{equation}\label{lumi_pgi}
P_\textrm{gI}(k, z, L) = A_\textrm{IA} b_1  P_{\delta \textrm{I}}(k, z) \left( \frac{1+z}{1+z_0} \right)^\eta \left( \frac{L}{L_0} \right)^\beta.
\end{equation}
Since the $z$-dependency has been shown to be negligible (\Singh; \citealt{johnston_kidsgama_2019};  \Sam), we fix $\eta=0$. 
In addition to this single-power-law model, a broken power law was introduced in \citet{fortuna_kids-1000_2021} to better account for the flattening of the $A_\textrm{IA}$ - $\log L$ relation at low luminosities. 

We test the single and double power law models, defined as:
\begin{align}
    A_\textrm{IA}(L) =& A_\beta \left\langle\frac{L}{L_0}\right\rangle^\beta ;
    \nonumber \\
    A_\textrm{IA}(L) =& A_\beta \left\langle\frac{L}{L_\textrm{break}}\right\rangle^\beta \
    \begin{cases} 
    \beta=\beta_1 & \text{if } L < L_\textrm{break}  \\
    \beta=\beta_2 & \text{if } L > L_\textrm{break} . \\
\end{cases}
\end{align}
Figure \ref{fig:lumi} demonstrates that both models provide decent fits to the data. The reduced $\chi^2$ for the single and double power-law are  $\chi^2$=2.42 with 28 degrees of freedom (30-2) and $\chi^2$=2.19 with 26 degrees of freedom (30-4), respectively. This indicates that both models perform rather poorly which is expected given the scatter in Fig.~\ref{fig:lumi}.
We can compute the BIC obtaining \begin{equation}
    \mathrm{Single \ PL: } \ \mathrm{BIC} = 74.6; \\
    \mathrm{Double \ PL: } \ \mathrm{BIC} = 70.6\\
\end{equation}
The difference $\Delta_{\mathrm{BIC}}(\mathrm{SPL}-\mathrm{DPL})=4.0$ indicates that the double power law is slightly favored, a result consistent with the recent literature. 

While converging on the $A_\textrm{IA}(L)$ relationship is important, we want to highlight the complexity that Fig.~\ref{fig:lumi} is omitting by using a single predictive parameter. While the redshift dependence has been found to be sufficiently well captured by the na\"{i}ve NLA model, many observational properties strongly affect the intrinsic alignment amplitude, as explained in the introduction. One can therefore not apply this intrinsic alignment-luminosity relationship to derive priors for a different survey without taking the risk of biasing the measurements. Including additional effects which explain the scatter in  Fig.~\ref{fig:lumi} will be the subject of future work. 

\begin{table}

\label{tab:lum}
\renewcommand{\arraystretch}{1.3}
\centering

\begin{tabular}{|c | c c |}   
\hline\hline       
 &  $\mathrm{log}\langle L/L_0 \rangle$  & $A_\textrm{IA}$\\ 
\hline                    
CMASS-UNIONS L1&  -0.26 & $2.74^{+0.59}_{-0.59}$  \\  
CMASS-UNIONS L2& -0.06 & $4.08^{+0.54}_{-0.54}$  \\
CMASS-UNIONS L3& 0.15 &$ 5.07^{+0.55}_{-0.55}$\\ 
CMASS-DES L1& -0.29 & $1.5^{+1.1}_{-1.1}$\\
CMASS-DES L2& -0.12 & $3.9^{+1.3}_{-1.3}$\\
CMASS-DES L3& 0.07 & $4.0^{+1.2}_{-1.2}$\\
   
\hline                  
\end{tabular}
\vspace{0.3cm}
\caption{Luminosity and intrinsic alignment amplitudes for the CMASS samples split into luminosity bins. Both CMASS-UNIONS and CMASS-DES show very strong scaling with luminosity.}
\end{table}

\subsection{Comparing UNIONS to other lensing samples}\label{sec:compa}

One of the main motivations of this work is a comparison of the measured intrinsic alignment quantities between UNIONS and DES Y3. The interest lies in the understanding of the universal nature of intrinsic alignment. As stated in the introduction, observational effects can influence this kind of measurement. Important effects discussed in the literature include the galaxy shape measurement method or the color of the sampled galaxies. Here we have a unique opportunity to study a homogeneous set of galaxies selected by the SDSS imaging survey, with two distinct shape measurement catalogs. While a number of similarities can be noted in the {\tt{ShapePipe}} and DES Y3 pipelines, a few important differences between the two datasets exist, as explained in Sect.~\ref{sec:cat}. A good agreement between the UNIONS and DES Y3 constraints would, therefore, indicate a certain control on the effects affecting intrinsic alignment measurements. This is crucial if one wants to use existing measurements as a prior for future surveys such as Rubin/LSST and \textit{Euclid}. Tight priors improve the statistical power to constrain cosmological parameters but will induce biases if inaccurate.

To test the compatibility of different shape measurement samples, we can compare our measurements to a third dataset for the CMASS-type galaxies  \citep{kurita_constraints_2023}. This dataset has the advantage of covering the entire CMASS sample and, therefore, overlaps with both UNIONS and DES. That study was performed with a newly developed, more accurate, 3D correlation function estimator \citep{kurita_power_2021}n, which decreases the error bars on the estimated parameters. Their measurement relies on the shape catalogue described in \citet{reyes_optical--virial_2012,mandelbaum_cosmological_2013}, which is based on the ``re-Gaussianization`` method described in \citet{hirata_shear_2003}. This method computes moments of the shape and corrects for a non-Gaussian PSF. Note that we make use of this shape catalogue for LOWZ shapes with which we validate our pipeline in App.~\ref{appen:lowz}.

\begin{figure*}
    \centering
    \hspace*{-0.5cm}
    \begin{minipage}{0.65\textwidth}
    \includegraphics[width=1.05\linewidth]{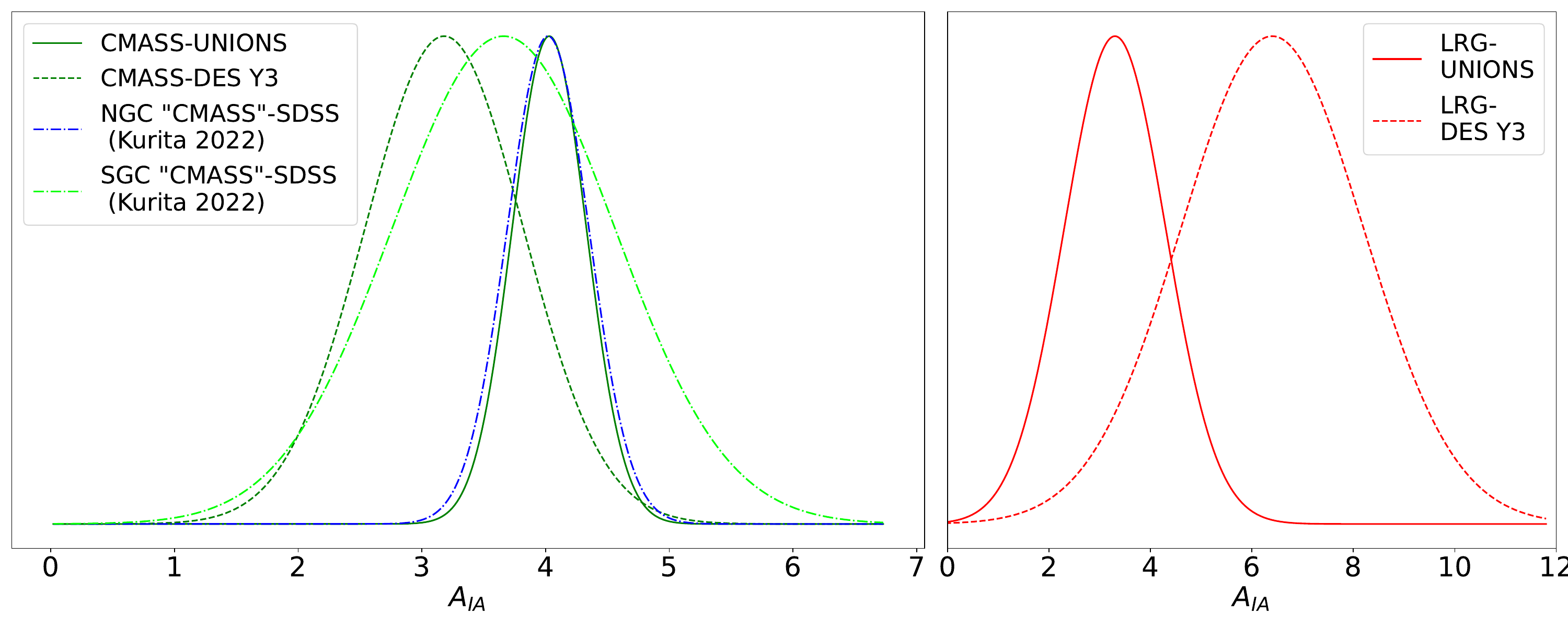}
    \end{minipage}
    \hspace*{1cm}
    \begin{minipage}{0.25\textwidth}
    \vspace*{-1cm}
    \caption{
    The 1D marginal posterior of the NLA intrinsic-alignment amplitude $A_\textrm{IA}$ of the same SDSS density tracer. \emph{Left panel:} UNIONS, DES Y3 (both our measurements), and the results from \citet{kurita_constraints_2023} in the north (NGC) and south galactic cap (SGC) using CMASS as density tracer. \emph{Right panel:} Our measured UNIONS and DES Y3 results with the LRG sample.}\label{fig:comp} 
    
    \end{minipage}

\end{figure*}

In Fig.~\ref{fig:comp} we show the overlap between the different posteriors. For the CMASS density tracer, the UNIONS and DES Y3 posteriors agree at $1.20\sigma$.
The samples do not differ in image selections but they differ slightly in redshift. The DES Y3 sample was chosen in the range $z\in[0.4,0.6]$, the UNIONS sample covers the range $z\in[0.4,0.7]$, and the Kurita sample is in the range $z\in[0.5,0.7]$. From previous studies on redshift dependence, we do not anticipate these discrepancies to result in significant variations. Although Fig.~\ref{fig:comp} might suggest a coherent north-south difference in intrinsic alignment, we do not consider this significant, given the inversion of the UNIONS and DES amplitudes in the LRG plot. Nonetheless, we report the differing mean galactic extinctions in Table 1, as this is the only spatially correlated property we could identify that might influence shape measurements.

For the LRG sample, we find an agreement between the DES and UNIONS measurements at the 1.51$\sigma$ level. We conclude that as long as the error bars are properly taken into account the measurements are in sufficiently good agreement to form a prior for stage IV surveys.

\subsection{Systematic diagnostics}

\begin{figure*}
\centering
\hspace*{-0.27cm}
\includegraphics[width=1.03 \textwidth]{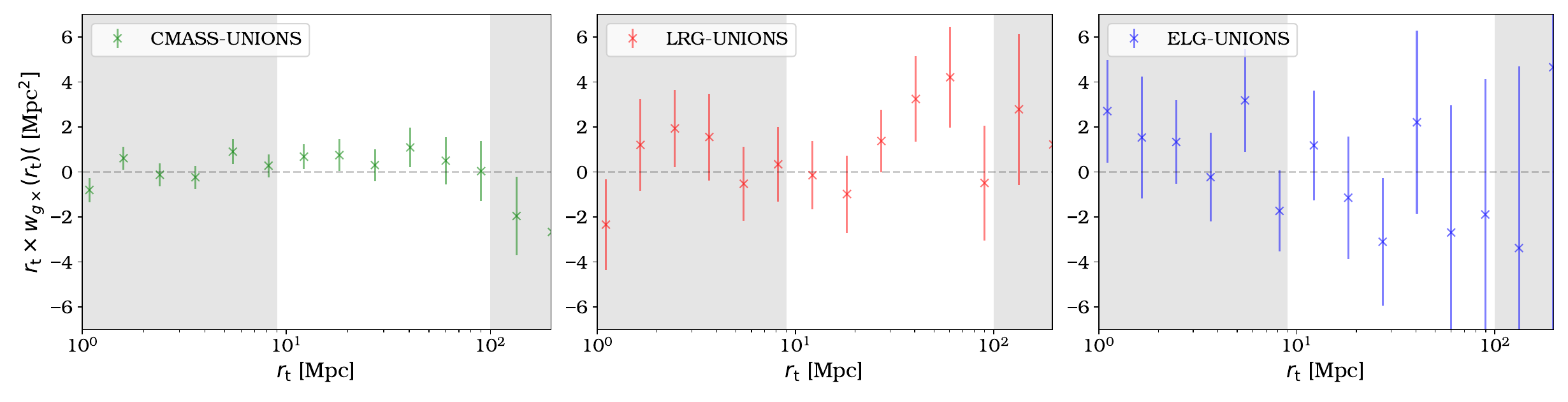}

\caption{The measurements of $w_{g \times}$ for the CMASS-UNIONS, LRG-UNIONS and ELG-UNIONS samples. Deviations from 0 are characterised in Table \ref{tab:wgx} and are all compatible at the $1\sigma$ level. }
\label{wgx}
\end{figure*}

In Fig.~\ref{wgx}, we show the cross-modes
$w_{\textrm{g}\times}$
computed from our samples. These are expected to be compatible with zero in the absence of systematic contributions \citep{joachimi_constraints_2011}, analogously to lensing B-modes in cosmic shear. In Table \ref{tab:wgx} we show the deviation from zero of the cross-mode $w_\textrm{g$\times$}$, and compare it to the one of the intrinsic-alignment mode $w_\textrm{g+}$. We convert the reduced $\chi^2$ into a Gaussian $\sigma$ estimate for easier interpretation. The cross-mode $w_\textrm{g$\times$}$ is compatible with zero at 1$\sigma$ for all samples, which strengthens the confidence in our results. In contrast, the deviations from zero for $w_{g+}$ testify of our high SNR measurement. For the ELG sample both $w_{g+}$ and $w_{g\times}$ remain consistent with zero at the 1$\sigma$ level.

\subsection{Quantifying PSF errors}

The effects of PSF mispecifications on cosmic shear have been studied in detail \citep{heymans_impact_2012,jarvis_dark_2020,giblin_kids-1000_2021}. Even small PSF residuals can produce biases in cosmological analyses and mimic the correlations induced by weak gravitational lensing. Quantities such as PSF leakage \citep{jarvis_science_2016}, $\rho$-statistics \citep{rowe_improving_2010}, and $\tau$-statistics \citep{gatti_dark_2021,guerrini_2024} have been introduced to measure and marginalize over biases. For direct measurements of intrinsic alignments, the impact of PSF errors has been less studied, as their effect is comparatively weaker. \citet{singh_intrinsic_2016} computed the additive bias from PSF mismodeling using the quantity

\begin{equation}
    A_\textrm{PSF}(r_\textrm{t} ) =
    \frac 1{
    r_\textrm{t,max} - r_\textrm{t,min}
    }
    \int\limits_{r_\textrm{t,min}}^{r_\textrm{t,max}} \textrm{d} r_\textrm{t}
    \frac{\langle \varepsilon_\textrm{obs} \varepsilon_\textrm{PSF} \rangle(r_\textrm{t})}{\langle \varepsilon_\textrm{PSF} \varepsilon_\textrm{PSF} \rangle(r_\textrm{t})}
    .
    \label{eq:A_PSF}
\end{equation}
This allows us to quantify a systematic contribution to the shape-shape correlation function $\xi_{++}$ given by
\begin{equation}
\xi_{++}^\mathrm{sys}\approx A^2_{PSF} \times \xi_{++}.
\end{equation}
A shape measurement method that does not correct for the PSF,
such as the isophotal method used in \cite{singh_intrinsic_2016}, 
can display a large amplitude of $A_\textrm{PSF}$.

Given the progress in imaging and shape measurement in recent years and the increasingly stringent requirements on additive biases, we calculate this quantity. We measure values of below $2\%$ for our red galaxies, which is sufficiently low to be ignored within our analysis. Interestingly, it varies between samples, which is the effect of a varying leakage across size and SNR, see also \citet{li_kids-legacy_2023}. In contrast to red galaxies, we find a $4\%$ additive bias for the ELG sample; such a bias can be calibrated via subtraction as in \citet{li_kids-legacy_2023}. We will address this in future work, as the current error bars on the ELG-UNIONS sample are large due to the limited number of galaxies.

\begin{figure}
    \centering
    \hspace*{-0.6cm}
    \includegraphics[width=1.08\linewidth]{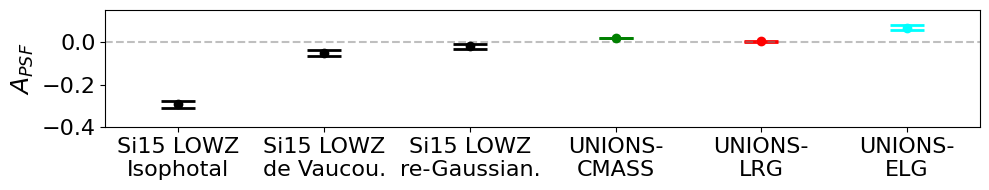}
    \caption{The amplitude Eq.~\eqref{eq:A_PSF} of the additive bias contamination of the shape-shape correlation function. From left to right, we show our measurements using the \Singh \ data corresponding to the isophotal, de Vaucouleurs, and re-Gaussianization shape catalogues, followed by UNIONS-CMASS, UNIONS-LRG, and UNIONS-ELG.
    }
    \label{fig:A_PSF}
\end{figure}

The measurement of $A_\textrm{PSF}$ informs us about additive leakage to the shape-shape correlation. Multiplicative and additive biases due to PSF contaminations for shape-density correlations have been recently quantified in \citet{zhang_point-spread_2024} [hereafter \ziwen]. Additive biases are not expected to contribute significantly, since the UNIONS PSF is not expected to show a dependence on the positions of SDSS galaxies. The so-called $\lambda$-statistics developed in \ziwen\ allow us to quantify the presence or absence of such biases. A multiplicative bias induced by the PSF can act as an additional calibration factor. 
In \ziwen \ $\lambda$-statistics were introduced for galaxy-galaxy lensing. Here, we adapt their work for the intrinsic-alignment correlation function $w_{\mathrm{g}+}$.

We start with the PSF error propagation model from \citet{paulin-henriksson_point_2008},
\begin{equation}
    \varepsilon^{\textrm{obs}} = \varepsilon^\textrm{int} + (1+m) \gamma + c + \delta \varepsilon + \alpha \varepsilon^\PSF.
    \label{eq:epsobs}
\end{equation}
In this model, the multiplicative bias term $m$ and additive bias term $c$ do not capture PSF information, which differs from the standard generic parametrization where $m$ and $c$ represent jointly all systematic contributions. The PSF leakage term $\alpha \varepsilon^\PSF$ captures a residual shape contribution from leakage of the PSF, typically attributed to an inaccurate PSF correction during the shape measurement process.
The $\delta \varepsilon$ term, on the other hand, captures residuals in the PSF measurement at the galaxy position and is due to errors in PSF measurement, modelling, and interpolation and can be understood as a PSF additive bias.

To estimate the PSF error at the galaxy position \citet{paulin-henriksson_point_2008} derived a formula using Gaussian error propagation,

\begin{equation}
    \delta \varepsilon
        = \left( \varepsilon^{\rm obs} - \varepsilon^\PSF \right)
        \frac{\delta T^\PSF}{T}
        - \frac{T^\PSF} T \delta \varepsilon^\PSF.
    \label{eq:delta_eps_original}
\end{equation}

The residuals on ellipticity $\delta \varepsilon^\PSF$ and size $\delta T^\PSF$ are obtained from a set of reserve stars, which are not used in the PSF modelling process. As we do not have this information at each position, we treat these residuals statistically as a constant prefactor.

With $w_{g+}$ we correlate intrinsic ellipticity with the position of a close-by galaxy (i.e. $d<\sqrt{\Pi_{\mathrm{max}}^2+r^2_t}$) via the estimator 
\begin{equation} \label{xi_p}
\xi_+=\langle \varepsilon_+ n\rangle.
\end{equation}
We denote with $\xi^{\mathrm{obs}}_+$ the observed correlation function and $\xi^{\mathrm{theo}}_+$ the underlying correlation of the intrinsic shape we are trying to quantify. 
We can define the change in correlation induced by a PSF misspecification as  
\begin{equation}\label{delta_xi}
    \delta \xi_+ = \xi_+^{\mathrm{obs}}-\xi^{\mathrm{theo}}
\end{equation}
By plugging Eq.~\eqref{eq:epsobs} into Eq.~\eqref{xi_p} 
we can write 
\begin{align}
    \delta \xi_+
        = \left\langle \varepsilon^\textrm{s}_\textrm{t} n \right\rangle +
        & \left\langle
            \frac{T^\PSF} T \frac{\delta T^\PSF}{T^\PSF} 
          \gamma_\textrm{t} n \right\rangle
        - \left\langle
            \frac{T^\PSF} T
            \frac{\delta T^\PSF} {T^\PSF} \varepsilon_\textrm{t}^\PSF n
          \right\rangle
        \nonumber \\
        & - \left\langle
            \frac{T^\PSF} T
            \delta \varepsilon_\textrm{t}^\PSF n
            \right\rangle
        + \left\langle \alpha
            \varepsilon_\textrm{t}^\PSF n
            \right\rangle.
\end{align}
The $\left\langle \varepsilon^\textrm{s}_\textrm{t} n \right\rangle$ contribution vanishes as intrinsic ellipticities do not correlate with random foreground positions
This allows us to define the $\lambda$-statistics,
\begin{align}
    \lambda_1
        & = \left\langle
                \varepsilon_\textrm{t}^\PSF \, n
            \right\rangle ;
    \nonumber \\
    \lambda_2
    & = \left\langle
            \frac{\delta T^\PSF} {T^\PSF} \varepsilon_\textrm{t}^\PSF \, n
        \right\rangle;
    \nonumber \\
    \lambda_3
    & = \left\langle
            \delta \varepsilon_\textrm{t}^\PSF \, n
        \right\rangle .
    \label{eq:lambda_123}
\end{align}
Since we do not expect the UNIONS PSF to be correlated with SDSS positions as stated above, the density tracer $n$ can here be considered as random positions. The $\lambda$-statistics then indicate additive errors due to random PSF errors, and potential PSF correlations with the mask or the survey footprint. 

\begin{figure}
    \centering
    \hspace*{-0.2cm}
    \includegraphics[width=1.\linewidth]{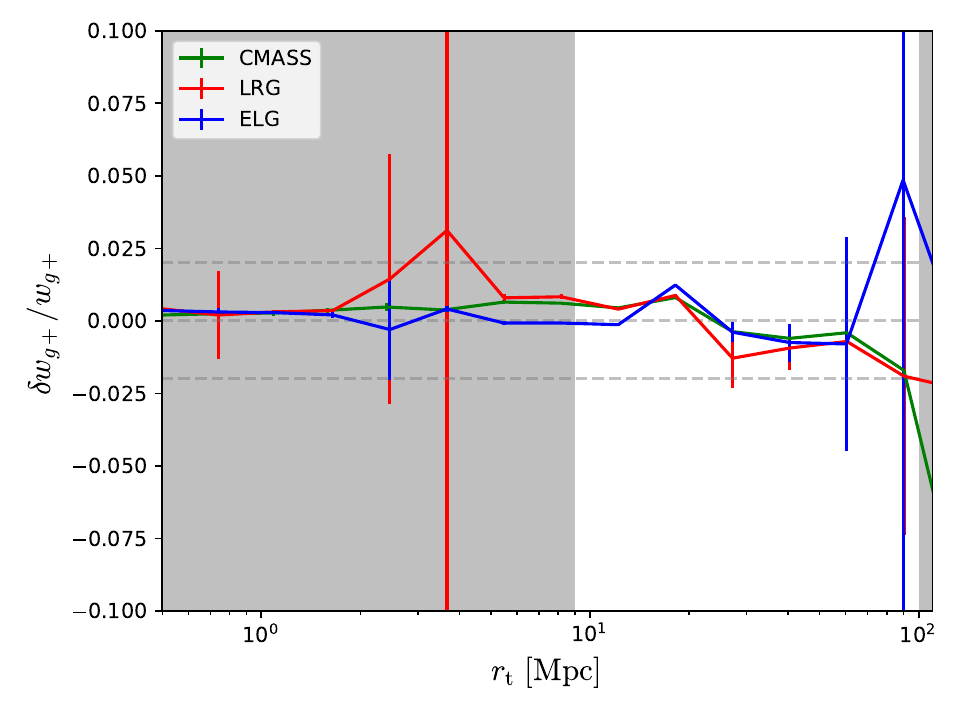}
    \caption{Relative contribution to the $w_{g+}$ signal of PSF induced corrrelations. We see that on the scales of interest the contribution is below $2\%$ which we judge satisfactory for current precision. The error bars blow up when $w_{g+}$ is close to 0. }
    \label{fig:lambda}
\end{figure}

The computation of $\lambda_1, \lambda_2, \lambda_3$, and $\alpha$ are detailed in \ziwen.
With that, we rewrite the PSF contamination to $\xi_+$ as
\begin{equation}
   \delta \xi_+ =
        \left\langle
            \frac{T^\PSF} T
        \right\rangle
        \left\langle
            \frac{\delta T^\PSF}{T^\PSF} 
        \right\rangle   \xi^{\mathrm{theo}}_+
     + \alpha \lambda_1
     - \left\langle
            \frac{T^\PSF} T
       \right\rangle
        \left(
            \lambda_2 + \lambda_3
        \right). 
    \label{eq:dg_lambda_theory}
\end{equation}
We can remove the dependence on the theoretical prediction $\xi^{\mathrm{theo}}_+$ by combining Eqs.~\eqref{delta_xi} and \eqref{eq:dg_lambda_theory} to write the bias of the measured quantity,
\begin{equation}
\label{eq:dg_lambda_obs}
   \delta\xi_+= \xi_+^{\rm obs} - \frac
   {
        \xi_+^{\rm obs}
        - \alpha \lambda_1
        +\left\langle
            \frac{T^\PSF} T
       \right\rangle
        \left(
            \lambda_2 + \lambda_3
        \right) \
   }
   {
        1 
        + \left\langle
            \frac{T^\PSF} T
        \right\rangle
        \left\langle
            \frac{\delta T^\PSF}{T^\PSF} 
        \right\rangle
    },
\end{equation}
To obtain the contamination to $w_{g+}$ we consider $\int_{\Pi_{\mathrm{min}}}^{\Pi\mathrm{max}}\delta \xi_+ \textrm{d}\Pi$. We plot $\delta w_{g+}/ w_{g+}$ for the CMASS, LRG and ELG samples in Fig.~\ref{fig:lambda}.
The PSF contamination remains under $2\%$ on the scales of interest.
This indicates that PSF errors are subdominant compared to our statistical power. In the future, when direct intrinsic alignment measurements reach precisions of a few percent, a significant measurement of PSF-tracer correlations can be used to account and calibrate PSF biases.

\begin{table}

\centering
\begin{tabular}{|c | c c c |}   
\hline      
 & CMASS- & LRG- & ELG  \\ 
 & UNIONS & UNIONS & UNIONS \\
\hline                    
   $\chi^2 \ [ w_{g\times} ] $& 0.65 (0.27$\sigma$) & 1.24  (0.86$\sigma$)  &  0.66 (0.27$\sigma$) \\  
   $\chi^2 \ [ w_{g+} ]  $& 11.7 (7.05$\sigma$)  & 1.89 (1.54$\sigma$)    & 0.95  (0.56$\sigma$) \\
   
\hline                  
\end{tabular}
\vspace{0.5cm}
\caption{Table testing the compatibility with the null hypothesis of the different samples. We indicate the significance with the reduced $\chi^2$ (6 d.o.f) and we convert it into a Gaussian standard deviation $\sigma$ deviation for interpretation. Note that these are deviations from 0, which are not equivalent to the significance of a given intrinsic alignment model.}        
\label{tab:wgx}

\end{table}

\section{Conclusion}
\label{sec:conclusion}

In this paper, we made use of the UNIONS shape catalogue and the large overlap with BOSS/eBOSS spectroscopy to carry out a direct measurement of the intrinsic alignment of galaxies. We obtain very significant measurements using  $201\, 639 $ CMASS, $78 \,134$ LRG, and $14 \,762$ ELG galaxies. Our results are consistent with previous studies showing a strong intrinsic alignment for red galaxies, and an alignment compatible with zero for emission line galaxies. These results demonstrate that the UNIONS observations with their excellent image quality are a competitive dataset to measure intrinsic alignment in the northern sky. In particular, we highlight the following points:
\begin{itemize}
    \item We obtain a detections of intrinsic alignment at $13\sigma$ with CMASS, at $3\sigma$ with LRG, and at $1\sigma$ with ELG. The latter is compatible with the null hypothesis. The NLA model was able to produce a compelling fit for these measurements in the range $r_\mathrm{t} \in [9,100] \, \mathrm{Mpc}$.
    \item We tested if the highest signal-to-noise measurement (CMASS-UNIONS) allows for a detection of the the higher-order terms of the TATT. For that, we fit transverse scales down to $4$ Mpc. We found no preference for non-zero values of these parameters. However, the data decorrelated the tidal field contribution $A_1$ from the density $\times$ tidal field contribution $b_{\mathrm{TA}}$. The bayesian information criteria seems to indicate a preference for the linear NLA model over TATT.
    \item Our measured NLA intrinsic-alignment amplitude scales with luminosity, as has been observed in the literature. A double power-law seems to provide a reasonable fit, although the reduced $\chi^2$ is large due to high scatter.  
    \item By adapting a novel framework from \ziwen, we estimated the bias contributions of the PSF leakage to the intrinsic alignment measurement, which is below 2\%.
    \item The cross-component $w_{\textrm{g}\times}$ was measured to be consistent with zero at $1\sigma$ for all samples, indicating an absence of obvious systematic contributions.
    \item 
    We re-analyze the DES Y3 data, and find intrinsic-alignment amplitude in agreement with UNIONS, at $1.2\sigma$ for CMASS and $1.5\sigma$ for LRGs.
    We further find agreement with other measurements obtained in \citet{kurita_constraints_2023} for CMASS galaxies. We conclude that all of these measurements are compatible with each other and can, therefore, serve as the basis for an informative prior for stage-IV surveys, as long as uncertainties are properly accounted for.
\end{itemize}
This work is a demonstration that the UNIONS sample is mature for competitive science cases. In the near future, we want to use the upcoming spectroscopic data in the northern hemisphere to probe intrinsic alignment on smaller scales and across a wider ranger of colors and luminosities. The goal is to improve our understanding at the dawn of stage-IV weak lensing surveys.

\begin{acknowledgements}
We thank Benjamin Joachimi and Elisa Chisari for informing discussions. FHP acknowledges support from the Centre National d'\'Etudes Spatiales (CNES). We are honored and grateful for the opportunity of observing the Universe from Maunakea and Haleakala, which both have cultural, historical and natural significance in Hawaii. This work is based on data obtained as part of the Canada-France Imaging Survey, a CFHT large program of the National Research Council of Canada and the French Centre National de la Recherche Scientifique. Based on observations obtained with MegaPrime/MegaCam, a joint project of CFHT and CEA Saclay, at the Canada-France-Hawaii Telescope (CFHT) which is operated by the National Research Council (NRC) of Canada, the Institut National des Science de l’Univers (INSU) of the Centre National de la Recherche Scientifique (CNRS) of France, and the University of Hawaii. This research used the facilities of the Canadian Astronomy Data Centre operated by the National Research Council of Canada with the support of the Canadian Space Agency. This research is based in part on data collected at Subaru Telescope, which is operated by the National Astronomical Observatory of Japan.
Pan-STARRS is a project of the Institute for Astronomy of the University of Hawaii, and is supported by the NASA SSO Near Earth Observation Program under grants 80NSSC18K0971, NNX14AM74G, NNX12AR65G, NNX13AQ47G, NNX08AR22G, 80NSSC21K1572 and by the State of Hawaii. LB is supported by the PRIN 2022 project EMC2 - Euclid Mission Cluster Cosmology: unlock the full cosmological utility of the Euclid photometric cluster catalog (code no. J53D23001620006).
\end{acknowledgements}

\bibliographystyle{aa}
\bibliography{my_zotero_manual,add_bib}
 \appendix

\section{Validating with LOWZ data from Singh 2015}\label{appen:lowz}
\begin{figure}\label{fig:lowz_wgp}
\includegraphics[width=1\linewidth]{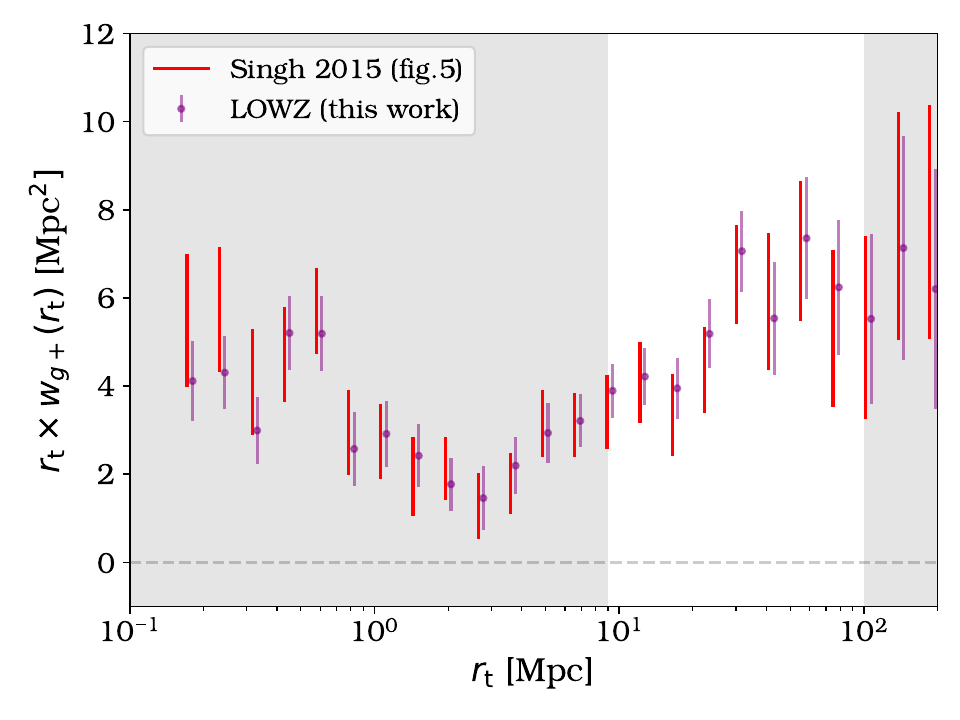}
\caption{Reproduction of the LOWZ measurement from \Singh \ for validation purposes. While we recovered a 2$\%$ agreement on the number of galaxies in the shape sample, the signals are in excellent agreement on all scales above 0.5 Mpc. At small scales, the number of pairs becomes very small, which makes the measurement noisier.  }
\end{figure}

To validate our pipeline our initial attempt was to use the \Sam \  DES measurements. During the course of this validation, we realized that their $ w_{g+}$ covariance matrix was incomplete as explained in Sect.~\ref{sec:cov}. To ensure good agreement with previous data sets we therefore validated our $w_{g+}$ measurement and covariance matrix estimate with data from \Singh. Due to some uncertainty in the way the extinction mask is applied, we only managed to reach a $2\%$ agreement on the number of galaxies in the shape catalog. Nevertheless, the agreement seen in Fig.\ref{fig:lowz_wgp} is visually very good. When fitting the NLA model we recover $A_1=4.62\pm0.44$,  in excellent agreement with the $A_1=4.6\pm0.5$ stated in the \Singh \ paper.

\begin{figure}\label{Lowz_cov}
\includegraphics[width=1.1\linewidth]{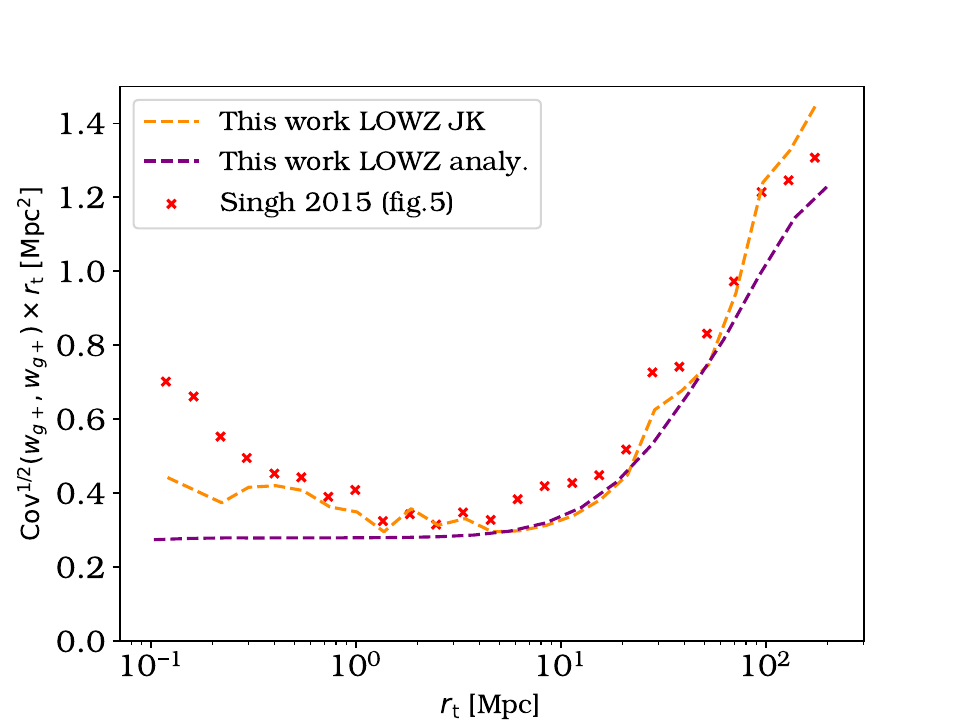}
\caption{We validate our covariance matrix estimate with the LOWZ sample from \cite{reyes_optical--virial_2012}, \cite{nakajima_photometric_2012}. We observe that our analytical prediction, the jackknife from Singh 2015 and our jackknife all agree very well, in particular on the scales of interest, while they deviate slightly below 1 Mpc.}
\end{figure}

To validate our $w_{gg}$ estimation we were satisfied by reaching agreement on $b_1$, $b_2$ with \Sam.  
We also use this opportunity to validate our analytical covariance matrix against the jackknife matrix estimates from LOWZ, we show the result in Fig.~\ref{Lowz_cov}. As can be expected from the correlation function plot, the agreement is good on all scales.

\section{Comparison to Lensfit}\label{appen:lensfit}
While the catalogue produced by {\tt{ShapePipe}} \citep{farrens_shapepipe_2022} is the main UNIONS catalogue, a second catalogue was produced by  {\it{lensfit}} to cross-check any systematics. For example it can be seen in \citet{li_kids-legacy_2023} that both agree very well when measuring lensing masses of AGN-black holes.
The strategy adopted in the {\it{lensfit}} pipeline is described in \citet{miller_bayesian_2013}. In short, the pipeline adopts a likelihood-based method by fitting a bulge+disk model in a Bayesian framework, which includes nuisance parameters, such as the galaxy position, which are marginalised over. 

\begin{figure}
    \centering
    \includegraphics[width=1.05
\linewidth]{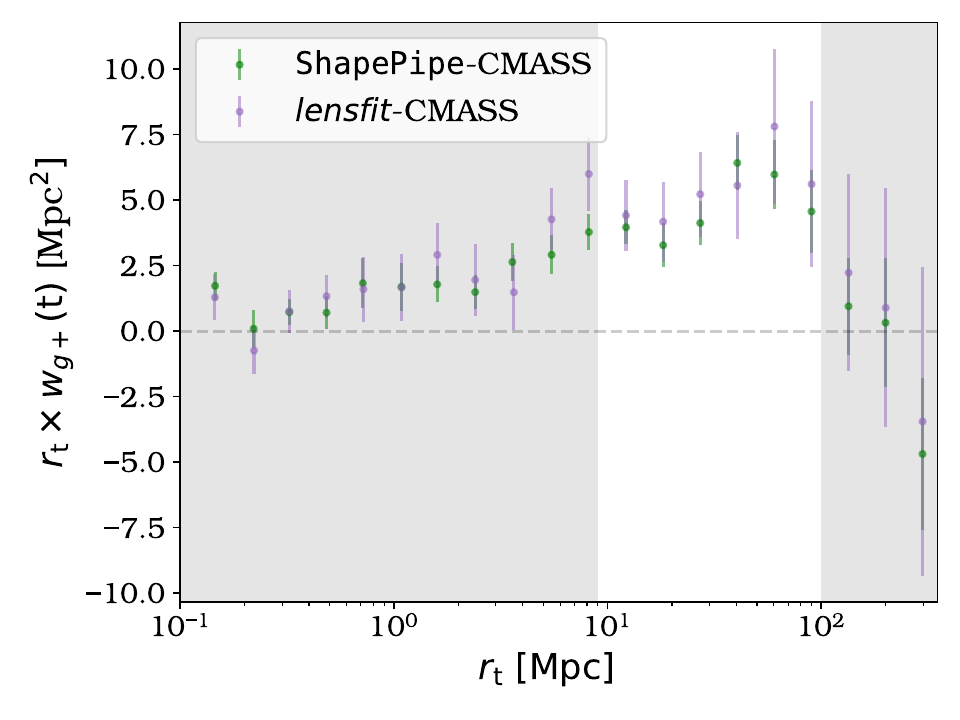}
    \caption{Comparison of {\it{lensfit}} and {\tt{ShapePipe}} estimates of $w_{g+}$. The good agreement increases confidence in the robustness of the measurement}
    \label{lensfit}
\end{figure}

While not used for the main results as the catalogue was not fully validated by the time of this work, we nevertheless verified that a good agreement exists as can be seen in Fig.\ref{lensfit}, where the CMASS-UNIONS data vectors generated with {\it{lensfit}} and {\tt{ShapePipe}} are in good agreement on all scales. While this is not a sufficient check to diagnose all systematics, it increases our confidence in our results. 

Note that it is not necessary that the $w_{\mathrm{g}+}$ from the two different shape methods agree, as they are based on very different model-fitting functions. This is known to potentially result in different intrinsic alignment amplitudes, as explained throughout this paper. While we initially believed that we would be able to obtain further constraints from the comparison of these two shape measurement samples, recent work by \cite{macmahon_intrinsic_2023} showed that a dedicated method was necessary to draw conclusions from a relative change in amplitude.
\end{document}